\documentclass[a4paper,12pt]{article}

\usepackage[utf8]{inputenc}
\usepackage{comment}
\usepackage[margin=2.5cm]{geometry}
\usepackage{graphicx}
\usepackage{caption}
\usepackage{subfig}
\usepackage{hyperref}
\usepackage{amsmath}
\usepackage{xcolor}
\usepackage{authblk}

\title{Quasinormal modes of rapidly rotating Ellis-Bronnikov wormholes}

\author[1]{Fech Scen Khoo\thanks{\href{mailto:fech.scen.khoo@uni-oldenburg.de}{fech.scen.khoo@uni-oldenburg.de}}}
\author[1]{Bahareh Azad\thanks{\href{mailto:bahareh.azad@uni-oldenburg.de}{bahareh.azad@uni-oldenburg.de}}}
\author[2]{Jose Luis Bl\'azquez-Salcedo\thanks{\href{mailto:jlblaz01@ucm.es}{jlblaz01@ucm.es}}} 
\author[2]{Luis Manuel Gonz\'alez-Romero\thanks{\href{mailto:mgromero@ucm.es}{mgromero@ucm.es}}} 
\author[1]{Burkhard Kleihaus\thanks{\href{mailto: b.kleihaus@uni-oldenburg.de}{b.kleihaus@uni-oldenburg.de}}} 
\author[1]{Jutta Kunz\thanks{\href{mailto:jutta.kunz@uni-oldenburg.de}{jutta.kunz@uni-oldenburg.de}}} 
\author[2]{Francisco Navarro-L\'erida\thanks{\href{mailto:fnavarro@fis.ucm.es}{fnavarro@fis.ucm.es}}} 
\affil[1]{Institut f\"ur  Physik, Universit\"at Oldenburg, Postfach 2503,
D-26111 Oldenburg, Germany}
\affil[2]{Departamento de F\'isica Te\'orica and IPARCOS, Facultad de Ciencias F\'isicas, Universidad Complutense de Madrid, Spain}

\date{\today}

\begin{document}

\maketitle

\begin{abstract}
We present for the first time a study of the quasinormal modes of rapidly rotating Ellis-Bronnikov wormholes in General Relativity. 
We compute the spectrum of the wormholes using a spectral decomposition of the metric perturbations on a numerical background.
We focus on the $M_z=2,3$ sector of the perturbations, and show that the triple isospectrality of the symmetric and static Ellis-Bronnikov wormhole is broken due to rotation, giving rise to a much richer spectrum than the spectrum of Kerr black holes. We do not find any instabilities for $M_z=2,3$ perturbations.
\end{abstract}

\section{Introduction}

Known only theoretically, wormholes represent fascinating solutions of gravitational theories (see e.g.~\cite{Morris:1988cz,Visser:1995cc,Lobo:2017}).
In General Relativity traversable wormholes can be obtained, by coupling it to a phantom field, yielding the well-known Ellis-Bronnikov wormholes \cite{Ellis:1973yv,Bronnikov:1973fh,Ellis:1979bh}.
Alternatively, the coupling to Dirac fermions also allows for traversable wormholes in General Relativity \cite{Blazquez-Salcedo:2020czn,Konoplya:2021hsm,Blazquez-Salcedo:2021udn}.

On the other hand, numerous alternative theories of gravity allow for classical traversable wormholes without the need for phantom fields, since additional terms in the gravitational field equations provide the necessary conditions \cite{Lobo:2017}.
Some specific examples are wormholes in Einstein-Gauss-Bonnet theory \cite{Kanti:2011jz,Kanti:2011yv,Antoniou:2019awm} and wormholes in beyond Horndeski theory \cite{Bakopoulos:2021liw}. 

By now much effort has been devoted to investigate the properties of wormholes, that might be observable. 
{Already since the '90s}, gravitational lensing by wormholes has been studied 
\cite{Cramer:1994qj,Safonova:2001vz,Perlick:2003vg,Nandi:2006ds,Abe:2010ap,Toki:2011zu,Nakajima:2012pu,Tsukamoto:2012xs,Kuhfittig:2013hva,Bambi:2013nla,Takahashi:2013jqa,Tsukamoto:2016zdu}.
In more recent years the shadows of wormholes have been investigated \cite{Bambi:2013nla,Nedkova:2013msa,Ohgami:2015nra,Shaikh:2018kfv,Gyulchev:2018fmd,Bouhmadi-Lopez:2021zwt,Guerrero:2022qkh,Huang:2023yqd}.
Moreover, the potential accretion disks of wormholes have been addressed and radiation associated with quasi-periodic oscillations has been predicted
\cite{Harko:2008vy,Harko:2009xf,Bambi:2013jda,Zhou:2016koy,Lamy:2018zvj,Deligianni:2021ecz,Deligianni:2021hwt}.

Of particular interest in many of these studies has been the question about how far wormholes can mimic the properties of black holes.
Indeed, under certain conditions wormholes can be almost perfect black hole mimickers \cite{Damour:2007ap}.
However, as shown in numerous examples (see references above) the potentially observable properties of wormholes can also be rather different from those of black holes.
For instance, the image of the wormhole and its environment seen in one universe (i.e., on one side of the throat) can be very different from the image seen in the other universe (i.e., on the other side of the throat) \cite{Huang:2023yqd}.

But not only observations in the electromagnetic spectrum might betray the presence of wormholes.
Ever since the successful detection of gravitational waves \cite{LIGOScientific:2016aoc}, the gravitational wave spectrum emitted by wormholes has also become a potential tool to distinguish wormholes from black holes.
We therefore here investigate the quasinormal modes of wormholes.
While for static wormholes the quasinormal modes have been discussed before \cite{Konoplya:2005et,Kim:2008zzj,Konoplya:2010kv,Konoplya:2016hmd,Volkel:2018hwb,Aneesh:2018hlp,Konoplya:2018ala,Blazquez-Salcedo:2018ipc,Konoplya:2019hml,Churilova:2019qph,Jusufi:2020mmy,
Bronnikov:2021liv,
Gonzalez:2022ote,Azad:2022qqn}, the quasinormal modes of rotating wormholes have not yet been addressed
(except for a study of the lowest scalar field modes in a Kerr-like spacetime \cite{Bueno:2017hyj}).

Since the treatment of quasinormal modes in the presence of rotation still poses a significant challenge, we here resort to a rather simple wormhole background, namely rotating Ellis-Bronnikov wormholes.
For slow rotation these wormhole solutions have been obtained analytically using perturbation theory \cite{Kashargin:2007mm,Kashargin:2008pk}, whereas for rapid rotation they are only known numerically \cite{Kleihaus:2014dla,Chew:2016epf}.
Note, that we are here only interested in asymptotically flat wormholes.
Certain non-asymptotically flat rotating wormholes are also known in closed form \cite{Cisterna:2023uqf}.

To construct the quasinormal modes we employ a spectral scheme.
This method captures a spectrum of modes, consisting of different $l$-led modes, including the fundamental as well as excitation modes for a given azimuthal number $M_z$.
Previously we have tested this scheme on the analytically known Kerr black hole solutions, reproducing the known Kerr quasinormal modes with high accuracy \cite{Blazquez-Salcedo:2023hwg}.
Here we present our first application of the method to an only numerically known background solution, the rapidly rotating Ellis-Bronnikov wormholes.

In section 2 we recall the rotating Ellis-Bronnikov wormhole solutions and some of their properties.
We present the metric and scalar field perturbations in section 3, introducing the perturbation functions, adopting suitable coordinates, and performing the spectral decomposition.
In section 4 we show our results for the quasinormal mode spectrum.
In particular, we show how rotation breaks the triple isospectrality of static symmetric Ellis-Bronnikov wormholes.
We end with our conclusions in section 5.

\section{Rotating Ellis-Bronnikov wormholes}
\label{setup}

\subsection{Theory and field equations}

We consider the theory of General Relativity coupled to a phantom scalar field $\varphi$,
\begin{eqnarray}
			S&=&\frac{1}{2\kappa}\int d^4x \sqrt{-g} 
		\Big[R + \frac{1}{2} \partial_\mu \varphi \, \partial^\mu \varphi 
		 \Big] \, .
   \label{eq:quadratic} 
\end{eqnarray}
The metric field equations are
\begin{eqnarray}
\mathcal{G}_{\mu\nu} = G_{\mu\nu} - T_{\mu\nu} = 0 \, ,
\end{eqnarray}
where $G_{\mu\nu}$ is the standard Einstein tensor and $T_{\mu\nu}$ is the phantom stress-energy tensor given by
\begin{eqnarray}
T_{\mu\nu} &=&  -\frac{1}{2}\partial_{\mu}\varphi\,\partial_{\nu}\varphi + \frac{1}{4}g_{\mu\nu}  (\partial \varphi)^2  \, .
\end{eqnarray}
For the phantom field, we have the field equation
\begin{eqnarray}
 \mathcal{S} =\frac{1}{\sqrt{-g}} \partial_{\mu} (\sqrt{-g} g^{\mu\nu} \partial_{\nu} \varphi)  = 0 \, .
\end{eqnarray}

\subsection{Stationary rapidly rotating background}

The {rapidly} rotating wormhole solutions in this theory were first constructed in \cite{Kleihaus:2014dla}.
There the following line element was considered,
\begin{eqnarray}
ds^2 &=& 
- \left[e^{f(r,\theta)}-e^{-f(r,\theta)}(r^2+r_0^2)w(r,\theta)^2\sin^2{(\theta)} \right] dt^2  
+ e^{\nu(r,\theta)-f(r,\theta)} dr^2  
 \nonumber \\
&+&
e^{\nu(r,\theta)-f(r,\theta)}(r^2+r_0^2) (d\theta^2 +\sin^2{(\theta)}d\phi^2) 
 \nonumber \\
&-& 2 e^{-f(r,\theta)}(r^2+r_0^2)w(r,\theta)\sin^2{(\theta)} dt d\phi
\, .
\label{metric_1}
\end{eqnarray}
It is parametrized by the functions 
$f$, $\nu$, and $w$ {that depend on the coordinates} $r$ and $\theta$.
The coordinate $r\in (-\infty,\infty)$ approaches two distinct Minkowski spacetimes. 
The variable $r_0$ 
is related to the equatorial radius of the throat $\mathrm{R}$, which is given by $\mathrm{R}=e^{-f(r=0, \theta=\pi/2)/2 }r_0$. 
The solutions are obtained numerically by solving the three
partial differential equations (PDEs) for $f$, $\nu$, and $w$, and the phantom scalar equation.

For $r \rightarrow +\infty$, the {asymptotic behaviour of} the background functions is
\begin{eqnarray}
    f(r,\theta) &=& -\frac{2M}{r} + \mathcal{O}(r^{-2}) \, ,  \\
    \nu(r,\theta) &=& \mathcal{O}
(r^{-2}) \, ,  \\
    w(r,\theta) &=& \frac{2J}{r^3} + \mathcal{O}(r^{-4})  \, ,
\end{eqnarray}
{while for $r \rightarrow -\infty$, it is}
\begin{eqnarray}
    f(r,\theta) &=& \frac{2M}{r} + \mathcal{O}(r^{-2}) \, ,  \\
    \nu(r,\theta) &=& \mathcal{O}(r^{-2}) \, ,  \\
    w(r,\theta) &=& w_{-\infty} + \frac{2J}{r^3} + \mathcal{O}(r^{-4}) \, . 
    \label{w_ninfty}
\end{eqnarray}
Here $M$ and $J$ are, respectively, the mass and the angular momentum of the wormholes. 
The constant $w_{-\infty}$ in equation (\ref{w_ninfty}) parameterizes different rotating solutions, where the static case is found for $w_{-\infty}=0$.

This constant ($w_{-\infty}$) is related to the angular velocity of the throat,  $\Omega=w(0,\theta)=w_{-\infty}/2$,
that defines the dimensionless rotational velocity of the throat,
\begin{eqnarray}
    \mathrm{v}_e = \mathrm{R}\,\Omega \, ,
\end{eqnarray}
where $\mathrm{R}$ is the equatorial radius of the throat.
In the limit of $\mathrm{v}_e \rightarrow 1$, the configuration approaches an extremal Kerr black hole \cite{Kleihaus:2014dla}. 

Another quantity of interest is the area of the throat, which is defined by the integral \cite{Chew:2016epf},
\begin{eqnarray}
    A_T = \int_0^{2\pi}\int_0^{\pi} r_0^2 e^{{\nu(0,\theta)/2-f(0,\theta)}}\sin{(\theta)}d\theta d\phi \, .
\end{eqnarray}
With this quantity we can characterize the size of the wormhole throat \cite{Kleihaus:2014dla, Chew:2016epf}, defining its areal radius,
\begin{eqnarray}
    \mathrm{R}_A = \sqrt{A_T/4\pi} \, .
    \label{areal_radius}
\end{eqnarray}

The set of solutions that we are going to study here
are symmetric solutions, such that the 
mass of the wormhole associated with the two infinities is the same. 
The non-symmetric generalization of the solutions can be found in
\cite{Chew:2016epf}.

The phantom scalar field is solved in closed form,
\begin{eqnarray}
    \varphi(r,\theta) = \frac{Q}{r_0}\left(\tan^{-1}{(r/r_0)}-\pi/2
    \right)\, ,
\end{eqnarray}
where $Q$ is the phantom scalar charge. 
The phantom field goes to zero at {plus infinity ($r \to +\infty$),
and to a finite constant at minus infinity ($r \to -\infty$).}

For further discussions of the geometry and properties of the wormhole solutions, 
we refer the readers, e.g., to \cite{Kleihaus:2014dla, Chew:2016epf}.

\section{Metric and scalar perturbations of rotating Ellis-Bronnikov wormholes}

After obtaining the background solutions,
we are all set to {study} the non-radial perturbations of the wormholes.
Recently we have developed a spectral method to study the general perturbations
of rotating geometries \cite{Blazquez-Salcedo:2023hwg}.
We have shown that our method works successfully for Kerr black holes.
But Kerr solutions are analytical solutions.
Here we present for the first time 
the application of our method to numerical solutions.
Most of the critical steps shown for Kerr black holes
in \cite{Blazquez-Salcedo:2023hwg} apply also
to Ellis-Bronnikov wormholes.

\subsection{Ansatz}
\label{sec_ansatz}

We perturb the background metric $g^{(bg)}_{\mu\nu}$,
given by equation (\ref{metric_1}),
to linear order in the perturbation parameter $\epsilon$,
\begin{eqnarray}
g_{\mu\nu} &=& g^{(bg)}_{\mu\nu} + \epsilon \delta h_{\mu\nu}(t,r,\theta,\phi) \, , \quad 
\end{eqnarray}
where $\delta h_{\mu\nu}$ is the linear perturbation of the metric,
\begin{eqnarray}
\delta h_{\mu\nu} &=& \delta h^{(A)}_{\mu\nu} + \delta h^{(P)}_{\mu\nu} \, . 
\, 
\end{eqnarray}
It {consists of} axial-led ($A$) and polar-led ($P$) perturbations. 
After fixing the gauge freedom, the axial-led metric perturbations are
\begin{equation}
 \delta h^{(A)}_{\mu\nu} = e^{i(M_z\phi-\omega t)} 
\begin{pmatrix}
0             & 0             & a_1(r,\theta) & a_2(r,\theta) \\
0             & 0             & a_3(r,\theta) & a_4(r,\theta) \\
a_1(r,\theta) & a_3(r,\theta) & 0             & 0 \\
a_2(r,\theta) & a_4(r,\theta) & 0             & 0
\end{pmatrix}   \, ,
\end{equation}
and
the polar-led metric perturbations are
\begin{equation}
 \delta h^{(P)}_{\mu\nu} = e^{i(M_z\phi-\omega t)} 
\begin{pmatrix}
N_0(r,\theta) & H_1(r,\theta) & 0             & 0 \\
H_1(r,\theta) & L_0(r,\theta) & 0             & 0 \\
0             & 0             & T_0(r,\theta) & 0  \\
0             & 0             & 0             & S_0(r,\theta) 
\end{pmatrix}   \, ,
\end{equation}
where
\begin{eqnarray}
a_1(r,\theta) &=& - i M_z \frac{h_0(r,\theta)}{\sin{\theta}} \, , \\
a_2(r,\theta) &=& \sin{\theta} \, \partial_\theta h_0(r,\theta) \, ,\\
a_3(r,\theta) &=& - i M_z \frac{h_1(r,\theta)}{\sin{\theta}} \, ,\\
a_4(r,\theta) &=& \sin{\theta} \, \partial_\theta h_1(r,\theta) \, ,\\
N_0(r,\theta) &=& \left( g^{(bg)}_{rr}(r,\theta) \right)^{-1}  N(r,\theta) \, ,\\
L_0(r,\theta) &=& \left( g^{(bg)}_{rr}(r,\theta) \right)  L(r,\theta) \, ,\\
T_0(r,\theta) &=& \left( g^{(bg)}_{\theta\theta}(r,\theta) \right) T(r,\theta) \, ,\\
S_0(r,\theta) &=& \left( g^{(bg)}_{\phi\phi}(r,\theta) \right) T(r,\theta)\, .
\end{eqnarray}

Apart from the metric, we have to linearly perturb the phantom field, meaning
\begin{eqnarray}
\varphi &=& \varphi^{(bg)} + \epsilon \delta\varphi(t,r,\theta,\phi) = \varphi^{(bg)} + \epsilon e^{i(M_z\phi-\omega t)} \Phi(r,\theta) \, .
\end{eqnarray}

We have factored out of the 
perturbations the
harmonic time dependence $e^{-i\omega t}$,
where $\omega$ is the frequency,
and the angle $\phi$-dependence $e^{iM_z\phi}$, where $M_z$ is the azimuthal number.

After perturbing the metric and the scalar field in the theory,
we arrive generically at
\begin{eqnarray}
\mathcal{G}_{\mu\nu} &=& \mathcal{G}_{\mu\nu}^{(bg)} + \epsilon \delta\mathcal{G}_{\mu\nu} e^{i(M_z\phi-\omega t)}  =0 \, , \\
\mathcal{S} &=& \mathcal{S}^{(bg)} + \epsilon \delta\mathcal{S} e^{i(M_z\phi-\omega t)}   =0 \, .
\end{eqnarray}
From the components of $\delta\mathcal{G}_{\mu\nu}$ and $\delta\mathcal{S}$ 
we have a
system of PDEs in $r$ and $\theta$ for the metric perturbation functions $N, L, T, H_1, h_1, h_0$ and the phantom
perturbation function $\Phi$. 
The coefficients of these perturbation {equations} depend on the background, 
the azimuthal number $M_z$ of the perturbations, and the frequency $\omega$. 
Note that this system of PDEs for rotating wormholes is more complicated than the one in \cite{Blazquez-Salcedo:2023hwg} for Kerr, since now the scalar perturbation is fully coupled to the metric perturbations.

\subsection{Parametrization, PDEs and boundary conditions}
\label{parametrization}

For computational convenience, we adopt the following coordinate {change} from $(r, \theta)$ to ($x, y$),
\begin{eqnarray}
    x = \frac{2}{\pi}\tan^{-1}
    {\left(\frac{r}{r_0}\right)} \, , \quad    y = \cos\theta \, .
\end{eqnarray}
The domain of integration is the square given by $-1 \le x \le 1$ and $-1 \le y \le 1$. 

We also parameterize the perturbation functions in terms of $x$ and $y$ as,
\begin{eqnarray}
 H_1 &=& \widetilde H_1(x,y) \frac{1}{(1-x)(1+x)^2} (1-y^2)^{M_z/2} e^{i \hat{R}} \, , 
 \label{eq_H1}
 \\
 T &=& \widetilde T(x,y) (1-y^2)^{M_z/2} e^{i \hat{R}} \, ,\\
 N &=& \widetilde N(x,y) \frac{1}{(1-x)(1+x)^2} (1-y^2)^{M_z/2} e^{i \hat{R}} \, ,\\
 L &=& \widetilde L(x,y) \frac{1}{1-x^2} (1-y^2)^{M_z/2} e^{i \hat{R}} \, ,\\
 h_0 &=& \widetilde h_0(x,y) \frac{1}{(1-x)(1+x)^2} (1-y^2)^{M_z/2} e^{i \hat{R}} \, ,\\
 h_1 &=& \widetilde h_1(x,y) \frac{1}{1-x^2} (1-y^2)^{M_z/2} e^{i \hat{R}} \, ,
 \label{eq_h1}\\
 \Phi &=& \widetilde P(x,y) (1-x^2) (1-y^2)^{M_z/2} e^{i \hat{R}}\, .
   \label{metric_param_ellis}
\end{eqnarray}
The tortoise coordinate $\hat{R}$ is chosen such that the perturbations are outgoing at both infinities. 
Note that the asymptotic behaviour of the functions $H_1$, $N$ and $h_0$ is different at both infinities. Even though flat 
spacetime is reached at $x=\pm 1$, in the coordinates we are using, only the background metric at $x=1$ has the standard Minkowski form, and hence the perturbation functions have the standard asymptotic form for outgoing waves. At $x=-1$, a coordinate transformation ($\phi \to \phi - w_{-\infty} t$) is necessary in order to bring the metric to the standard Minkowski form.
 We choose the following components of the linearized (Einstein) equations for numerical integration:
 $\{ \delta\mathcal{G}_{tr} \, , \delta\mathcal{G}_{t\theta} \,,  \delta\mathcal{G}_{rr} \,,  \delta\mathcal{G}_{r\theta} \,,  \delta\mathcal{G}_{r\phi} \,,  \delta\mathcal{G}_{\theta\phi} \,,  \delta\mathcal{S}\}$.
 The remaining four linearized Einstein equations,
 $\{ \delta\mathcal{G}_{tt}, \delta\mathcal{G}_{t\phi},  \delta\mathcal{G}_{\theta\theta},  \delta\mathcal{G}_{\phi\phi} \}$,
 are used to test the accuracy of the resulting quasinormal modes.

Therefore the system of PDEs in $(x,y)$ that we need to solve can be written in operator form as
\begin{eqnarray}
    \mathcal{D}_{\mathrm{I}}(x,y) \vec{X}(x,y) = 0 \,, \qquad  {\mathrm{I}} = 1,...,7 \, ,
    \label{metric_eq_xy_ellis}
\end{eqnarray}
where $\vec{X}=[H_1, T, N, L, h_0, h_1, \Phi]$ is a vector consisting of all the perturbation functions (in general complex) and $\mathcal{D}_{\mathrm{I}}$ are seven linear operators that depend non-trivially on the background solution.

Equation (\ref{metric_eq_xy_ellis}) has to be satisfied in the interior of the domain of integration, while on the boundaries we impose the appropriate boundary conditions for the perturbations.

Requiring an outgoing wave solution at plus infinity (i.e.~the boundary at $x=1$)  
 implies that the perturbation functions behave as
\begin{eqnarray}
           T &=& e^{i\omega R^* } \left( T^{+}(\theta) + \mathcal{O}\left(\frac{1}{r}\right) \right) \, ,\\
           H_{1} &=& r e^{i\omega R^* } \left( H^{+}_{1}(\theta) + \mathcal{O}\left(\frac{1}{r}\right) \right)\, ,\\
           N &=& r e^{i\omega R^* } \left( N^{+}(\theta) + \mathcal{O}\left(\frac{1}{r}\right)  \right) \, ,\\
           L &=& r e^{i\omega R^* } \left( L^{+}(\theta) + \mathcal{O}\left(\frac{1}{r}\right)  \right) \, ,\\
           h_{0} &=&  r e^{i\omega R^* } \left( h^{+}_{0}(\theta) + \mathcal{O}\left(\frac{1}{r}\right) \right)\, ,\\
           h_{1} &=&  r e^{i\omega R^* } \left( h^{+}_{1}(\theta)  + \mathcal{O}\left(\frac{1}{r}\right) \right)\, ,\\
           \Phi &=& \frac{1}{r} e^{i\omega R^* } \left(  \Phi^{+}(\theta) + \mathcal{O}\left(\frac{1}{r}\right)   \right)\, ,
\end{eqnarray}
expressed in $(r, \theta)$ variables for clarity, and $ \frac{dR^*}{dr} =  1 + \frac{2M}{r} + \mathcal{O}\left(\frac{1}{r^2}\right)  $.
Introducing these expansions in the perturbation equations
results in
a set of seven conditions that can be written in operator form as
\begin{eqnarray}
    \mathcal{A}_{\mathrm{I}} (x,y) \vec{X}(x,y)|_{x=1} = 0 \,, \quad  {\mathrm{I}} = 1,...,7 \, ,
\label{bcg_inf_ellis}
\end{eqnarray}
where $\mathcal{A}_{\mathrm{I}}$ are linear operators.
Similarly, imposing an outgoing wave solution at minus infinity ($x=-1$)
 implies the following 
behaviours,
\begin{eqnarray}
           T &=& e^{-i(\omega-M_z w_{-\infty}) R^* } \left( T^{-}(\theta) + \mathcal{O}\left(\frac{1}{r}\right) \right) \, ,\\
           H_{1} &=& r^2 e^{-i(\omega-M_z w_{-\infty}) R^* } \left( H^{-}_{1}(\theta) + \mathcal{O}\left(\frac{1}{r}\right) \right)\, ,\\
           N &=& r^2 e^{-i(\omega-M_z w_{-\infty}) R^* } \left( N^{-}(\theta) + \mathcal{O}\left(\frac{1}{r}\right)  \right) \, ,\\
           L &=& r e^{-i(\omega-M_z w_{-\infty}) R^* } \left( L^{-}(\theta) + \mathcal{O}\left(\frac{1}{r}\right)  \right) \, ,\\
           h_{0} &=&  r^2 e^{-i(\omega-M_z w_{-\infty}) R^* } \left( h^{-}_{0}(\theta) + \mathcal{O}\left(\frac{1}{r}\right) \right)\, ,\\
           h_{1} &=&  r e^{-i(\omega-M_z w_{-\infty}) R^* } \left( h^{-}_{1}(\theta)  + \mathcal{O}\left(\frac{1}{r}\right) \right)\, ,\\
           \Phi &=& \frac{1}{r} e^{-i(\omega-M_z w_{-\infty}) R^* } \left(  \Phi^{-}(\theta) + \mathcal{O}\left(\frac{1}{r}\right)   \right)\, ,
\end{eqnarray}
expressed again in $(r, \theta)$ variables, and $ \frac{dR^*}{dr} =  1 - \frac{2M}{r} + \mathcal{O}\left(\frac{1}{r^2}\right)  $.
This results in
\begin{eqnarray}
    \mathcal{B}_{\mathrm{I}}(x,y) \vec{X}(x,y)|_{x=-1} = 0 \,, \quad  {\mathrm{I}} = 1,...,7 \, ,
\label{bcg_ninf_ellis}
\end{eqnarray}
with other linear operators $ \mathcal{B}_{\mathrm{I}}$.

In addition we have to require regularity of the perturbation functions on the {axis of rotation}. At the upper part of the axis ($y=1$), this results in a set of seven conditions,
\begin{eqnarray}
    \mathbf{\alpha}_{\mathrm{I}}(x,y) \vec{X}(x,y)|_{y=1} = 0 \,, 
     \quad \mathrm{I} = 1,...,7 \, ,
\label{bcg_np}
\end{eqnarray}
where we used the expansions
\begin{eqnarray}
           T &=& T^{NP}(x)  + \mathcal{O}\left(y-1\right)   \, ,\\
           H_{1} &=& H_1^{NP}(x)  + \mathcal{O}\left(y-1\right)   \, ,\\
           N &=& N^{NP}(x)  + \mathcal{O}\left(y-1\right)   \, ,\\
           L &=& L^{NP}(x)  + \mathcal{O}\left(y-1\right)   \, ,\\
           h_0 &=& h_0^{NP}(x)  + \mathcal{O}\left(y-1\right)   \, ,\\
           h_1 &=&  h_1^{NP}(x)  + \mathcal{O}\left(y-1\right)   \, ,\\
           \Phi &=& \Phi^{NP}(x)  + \mathcal{O}\left(y-1\right)   \, .
\end{eqnarray}
Similarly at the lower part of the axis $(y=-1)$ we find,
\begin{eqnarray}
    \mathbf{\beta}_{\mathrm{I}}(x,y) \vec{X}(x,y)|_{y=-1} = 0 \,, 
     \quad \mathrm{I} = 1,...,7 \, ,
\label{bcg_sp}
\end{eqnarray}
where the expansions used are
\begin{eqnarray}
           T &=& T^{SP}(x)  + \mathcal{O}\left(y+1\right)   \, ,\\
           H_{1} &=& H_1^{SP}(x)  + \mathcal{O}\left(y+1\right)   \, ,\\
           N &=& N^{SP}(x)  + \mathcal{O}\left(y+1\right)   \, ,\\
           L &=& L^{SP}(x)  + \mathcal{O}\left(y+1\right)   \, ,\\
           h_0 &=& h_0^{SP}(x)  + \mathcal{O}\left(y+1\right)   \, ,\\
           h_1 &=&  h_1^{SP}(x)  + \mathcal{O}\left(y+1\right)   \, ,\\
           \Phi &=& \Phi^{SP}(x)  + \mathcal{O}\left(y+1\right)   \, .
\end{eqnarray}

\subsection{Spectral decomposition}

In order to solve the previous system of PDEs and boundary conditions, we follow the same procedure we used in \cite{Blazquez-Salcedo:2023hwg} for Kerr. 

First we decompose the radial part of the perturbations in Chebyshev polynomials of the first kind, $T_{k}(x)$, and the angular part of the perturbations in Legendre polynomials of the first kind, $P_l^{M_z}(y)$.
Thus the perturbation functions have the following expression
\begin{eqnarray}
     \widetilde X_{\mathrm{I}}(x,y) &=& \sum_{k=0}^{N_x-1}  \, \, \sum_{l=|M_z|}^{N_y+|M_z|-1} C_{{\mathrm{I}},k,l}  \,  T_{k}(x)  \,  P_l^{M_z}(y)  \,  (1-y^2)^{-M_z/2} \, ,
\end{eqnarray}
for ${\mathrm{I}}=1,...,7$, i.e.~the seven perturbation functions that form the PDE system. The parameters $N_x$ and $N_y$ are the number of grid points chosen in the direction of the $x$ and $y$ coordinates, respectively.
Each expansion of the perturbation functions
comes with a set of constants $C_{{\mathrm{I}},k,l}$ that we have to determine from the system of PDEs and boundary conditions. Note that there are a total of $7 \times N_x \times N_y$ such constants.

Finally, we discretize the domain of integration for a grid of size $N_x \times N_y$.
In the $x$ direction we choose the Gauss-Lobato points,
\begin{eqnarray}
x_I = \cos{\left(\frac{I-1}{N_x-1}\pi\right)} \, , \quad  I=1, ..., N_x \, , 
\end{eqnarray}
and in $y$ direction a homogeneous mesh, 
\begin{eqnarray}
   y_K = 2\frac{K-1}{N_y-1}-1 
   \, ,  \quad K=1,...,N_y \, . 
\end{eqnarray}
We evaluate the seven perturbation equations at each grid point. Recall, that the perturbation equations depend on the background metric, which for the rapidly rotating wormholes that we are considering is known only in numerical form. Hence in order to evaluate the seven perturbation equations, we need to calculate numerically the values of the functions $f$, $\nu$ and $w$ and the corresponding partial derivatives at each point of the grid. This is done using interpolations with sixth order splines\footnote{We check that the interpolated background functions satisfy the background PDEs with an accuracy of at least $10^{-8}$.}.

After evaluating the PDEs and boundary conditions on each point of the grid, we are left with a total of $7 \times N_x \times N_y$ algebraic equations for the constants $C_{{\mathrm{I}},k,l}$. This system of algebraic equations can be written in a matrix form,
\begin{eqnarray}
    \left( \mathcal{M}_0 + \mathcal{M}_1 \omega + \mathcal{M}_2 \omega^2 \right) \Vec{C} = 0 \, ,
    \label{matrix_eq}
\end{eqnarray}
where $\vec{C}$ is a vector whose components are all the constants $C_{{\mathrm{I}},k,l}$ and $\omega$ is the complex frequency.
The square matrices
$\mathcal{M}_0$, $\mathcal{M}_1$ and $\mathcal{M}_2$ are numerical matrices, each of size $(7 \times N_x \times N_y) \times (7 \times N_x \times N_y)$. 
This is essentially a quadratic eigenvalue problem (\ref{matrix_eq}). 
In order to determine the quasinormal modes, we implement our numerical procedures using Maple and Matlab with the Multiprecision Computing Toolbox Advanpix \cite{Advanpix}.

To check the validity of our method, 
we have first reproduced the quasinormal mode spectrum 
in the static configuration
found in \cite{Blazquez-Salcedo:2018ipc, Azad:2022qqn} which employ a different method (shooting method). 
We find that with values of $N_x$ and $N_y$ typically of 20, we get 
an accuracy of $10^{-6}$ in these static modes.
Furthermore, we have also reproduced the quasinormal mode spectrum for Kerr black holes
with an accuracy of $10^{-6}$ using our method with
$N_x=N_y=20$
\cite{Blazquez-Salcedo:2023hwg}.

\section{Spectrum of quasinormal modes for rotating Ellis-Bronnikov wormholes}

In this section we discuss the results we obtain by applying our spectral method described in the previous sections. We will focus on the azimuthal number $M_z=2,3$ sector of the perturbations, which is the one that is typically expected to dominate the gravitational wave emission during the ringdown phase. Other values of $M_z$ will be studied elsewhere.

\begin{figure}
    \centering
    \includegraphics[angle=-90,width=0.8\textwidth]{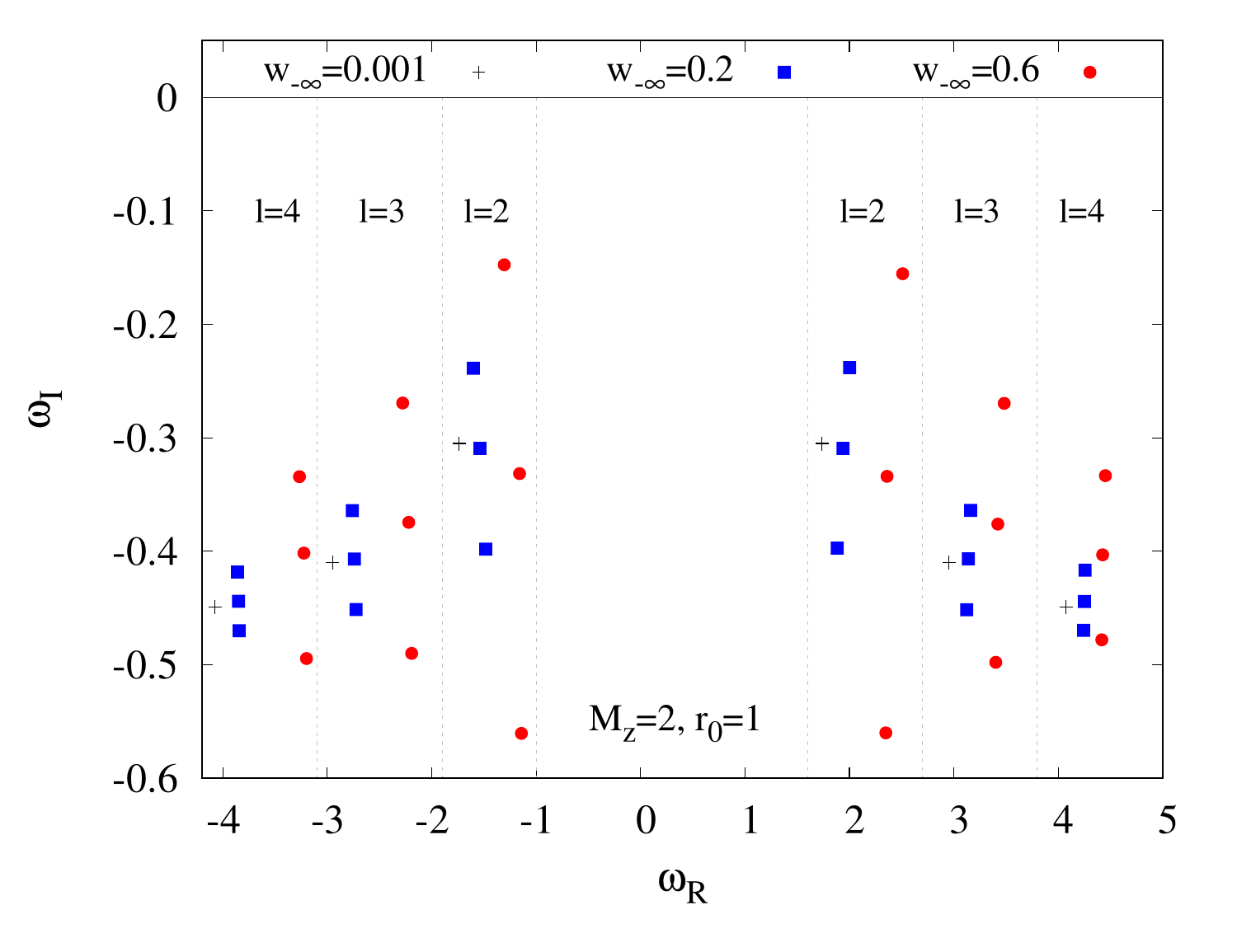}
    
    \caption{An overview of the $M_z=2$ co-rotating and counter-rotating modes for three rotating Ellis-Bronnikov solutions with $r_0=1$ and $w_{-\infty}=0.001$ (black crosses), $w_{-\infty}=0.2$ (blue squares) and $w_{-\infty}=0.6$ (red dots). The vertical lines are added as a reference for our discussions about the leading multipolar behaviour of the modes.}
    \label{fig:spectrum_sample}
\end{figure}

Figure \ref{fig:spectrum_sample} shows a typical spectrum we can obtain with our method.
It contains the fundamental modes for several multipoles $l$.
In the figure we show the eigenvalue $\omega$ in the complex plane for $M_z=2$ and $r_0=1$. 
In black crosses we show the quasinormal modes for an almost static Ellis-Bronnikov solution with $w_{-\infty}=0.001$. The blue squares represent the quasinormal modes from a solution with $w_{-\infty}=0.2$. In red dots we show the modes from a solution with $w_{-\infty}=0.6$. 
{We typically choose grids with $N_x=N_y=20$, for which all the} PDEs are satisfied with an accuracy of at least $10^{-4}$ at every point of the grid.
These parameters have been shown to work satisfactorily in reproducing the mode spectrum for the static Ellis-Bronnikov wormholes and Kerr black holes.

Let us first focus on the modes for the almost static solution (black crosses). In the region that we are plotting we can distinguish six modes, three with positive values of $\omega_R$ and three with negative values of $\omega_R$. These are the modes led by $l=2,3,4$ multipoles in the angular dependence. The ones closest to the $\omega_R=0$ axis are the fundamental $l=2$ modes, and as we move away from this axis we find the $l=3$ and $l=4$ modes. As a reference we have added the vertical lines that roughly indicate the regions where the $l=2,3,4$ modes reside.

In our previous work we showed that the static and symmetric Ellis-Bronnikov solution possesses a triple isospectrality between the axial, polar and scalar modes \cite{Azad:2022qqn}. 
This means that 
in Figure \ref{fig:spectrum_sample} 
at each black cross (almost static Ellis-Bronnikov), there are three different solutions to the perturbation functions with almost indistinguishable values of $\omega$.

As we spin up the background, the spectrum becomes much richer as the 
solutions are no longer degenerate. 
For instance, consider the modes for the solution with $w_{-\infty}=0.2$ (blue squares). 
While we have a single degenerate mode in the static limit (black crosses), there are now three distinct modes. 

Let us for instance take a look at the range of $1.6<\omega_R<2.7$ 
in Figure \ref{fig:spectrum_sample}
that roughly indicates the $l=2$ region.
We can see that one of the modes is still close to the almost static mode, but the other two modes are much more sensitive to the angular momentum of the solution. The imaginary part of one of them increases significantly, while for the other one it decreases. 

When the background rotates, it is no longer possible to decouple the perturbations using spherical harmonics. 
Nonetheless it is possible, as we will explicitly show later, to characterize the modes by their leading multipolar behaviour. 
Hence, following the notion we introduced in \cite{Blazquez-Salcedo:2023hwg}, we refer to the modes in the $l=2$ region as the $(l=2)$-led modes, etc.

When we increase the angular momentum of the background even further (e.g.~the red dots with $w_{-\infty}=0.6$), the separation between the three modes increases as well. 
At $w_{-\infty}=0.6$ it is still possible to identify the regions of different $l$ with vertical lines.
For faster spinning background these regions could overlap. 
Hence, in order to distinguish the nature of the modes, it is convenient to study the spectrum by increasing the angular momentum of the background gradually. 
This way we can track the modes continuously from their static limit as we spin up the solution.

Figure \ref{fig:spectrum_sample} shows that with our method we can obtain the modes with positive and negative values of $\omega_R$. 
In fact we found the following {empirical} relation between the co-rotating modes $\omega_+$ (with $\omega_R>0$) and the counter-rotating modes $\omega_-$ (with $\omega_R<0$),
\begin{equation}
    \omega_- = -\omega_+^* + M_z w_{-\infty} \, .
\end{equation}
This relation is satisfied within the numerical accuracy of our calculations. 
Therefore without loss of generality, we will focus only on the co-rotating modes  $\omega_R>0$ in the following.

\begin{figure}
    \centering
    \includegraphics[angle=-90,width=0.8\textwidth]{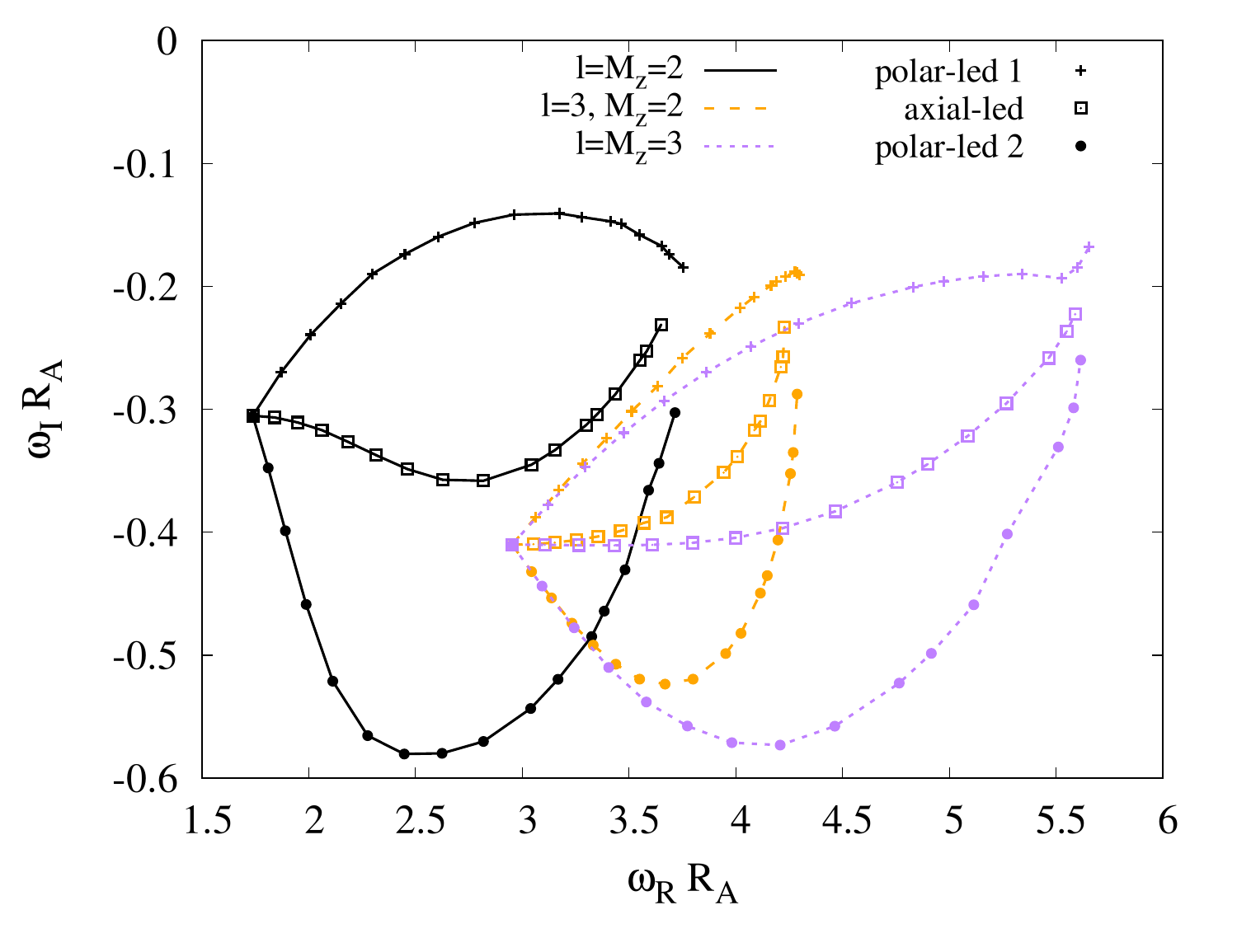}
    
    \caption{The imaginary part of $\omega$ as a function of the real part of $\omega$, both scaled with the areal radius $\mathrm{R}_A$. In black the $l=M_z=2$ modes, in orange the $l=3,M_z=2$ modes, and in purple the $l=M_z=3$ modes. In crosses we show the primary polar-led branch, in squares the axial-led branch, and in dots the secondary polar-led branch. The rotational velocity of the background solution considered is in the range of $0<\mathrm{v}_e<0.82$.
    We provide the values of the modes in Tables \ref{tab:l2m2}, \ref{tab:l3m2}, and
    \ref{tab:l3m3}.
    }
    \label{fig:ellis_qnms_wrwi}
\end{figure}

In Figure \ref{fig:ellis_qnms_wrwi} we show our results as we track the modes by taking small increments in the angular momentum of the wormhole. 
Here we scale the complex $\omega$ with the areal radius of the wormhole throat, $\mathrm{R}_A$ (\ref{areal_radius}). In black we show the fundamental $(l=2)$-led modes with $M_z=2$ perturbations. For simplicity we will refer to these as the $l=M_z=2$ modes. 
Similarly, in orange we show the $l=3, M_z=2$ modes, and in purple the $l=M_z=3$ modes. 
From the figure, we see that the real part of the frequencies grows as we increase the angular momentum, while the imaginary part increases or decreases depending on the nature of the mode.

We can characterize each mode by the nature of the leading perturbation functions, i.e., whether the polar or the axial components dominate. In Figure \ref{fig:ellis_qnms_wrwi} we show in crosses the primary branch of polar-led modes (denoted by polar-led 1). 
These are modes for which the polar functions prevail over the axial ones, 
and they possess the longest damping time. In dots we show the secondary branch of polar-led modes (denoted by polar-led 2), which have shorter damping times. 
In between these two families of polar-led modes, we always find the axial-led modes, i.e., modes for which the axial functions prevail over the polar ones. 
Note that as we increase the spin of the background, there seems to be a tendency for the modes to come back together. We will comment more on this later.

\begin{figure}
    \centering
    \includegraphics[trim=50 0 50 50,clip,angle=-90,width=0.32\textwidth]{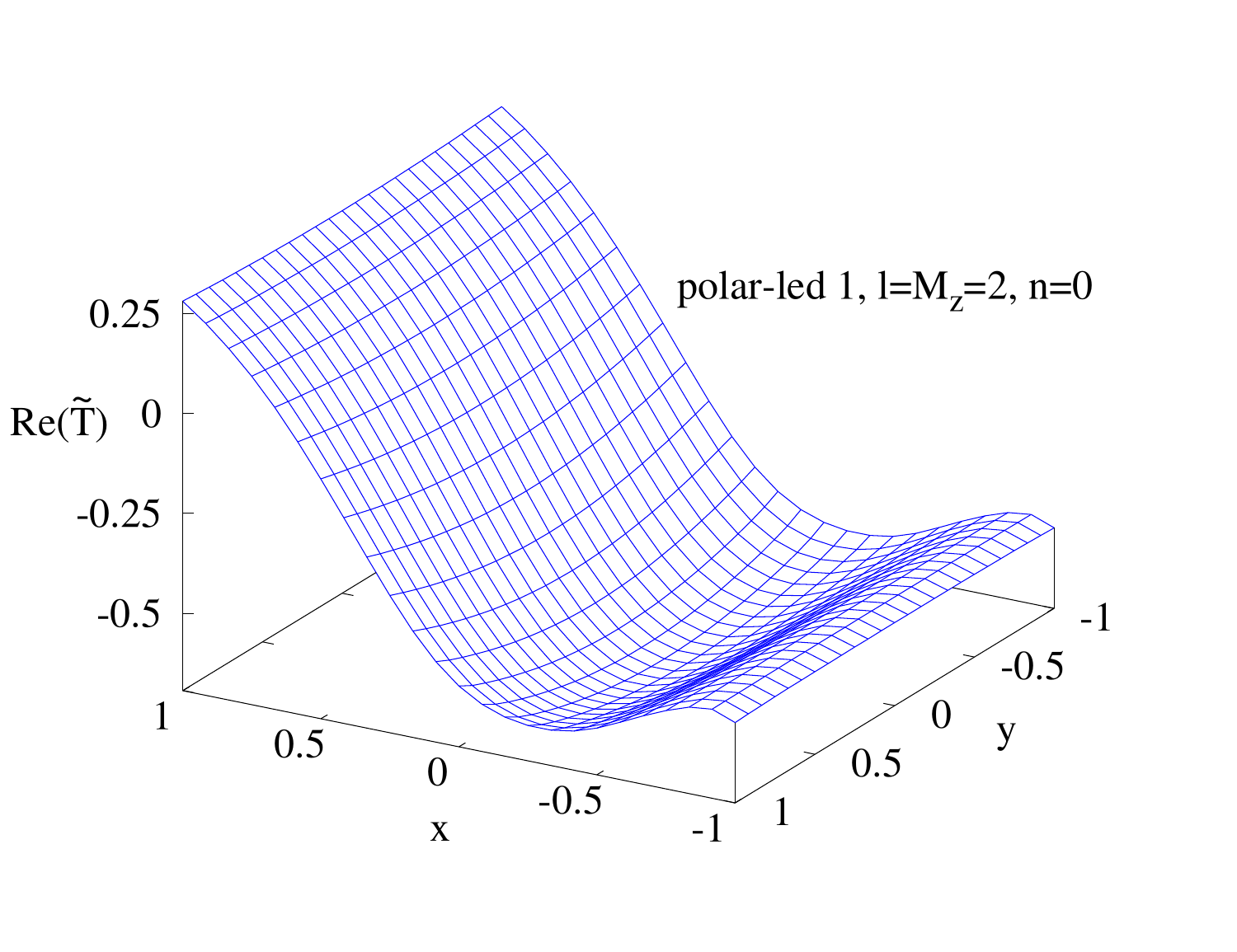}
    \includegraphics[trim=50 0 50 50,clip,angle=-90,width=0.32\textwidth]{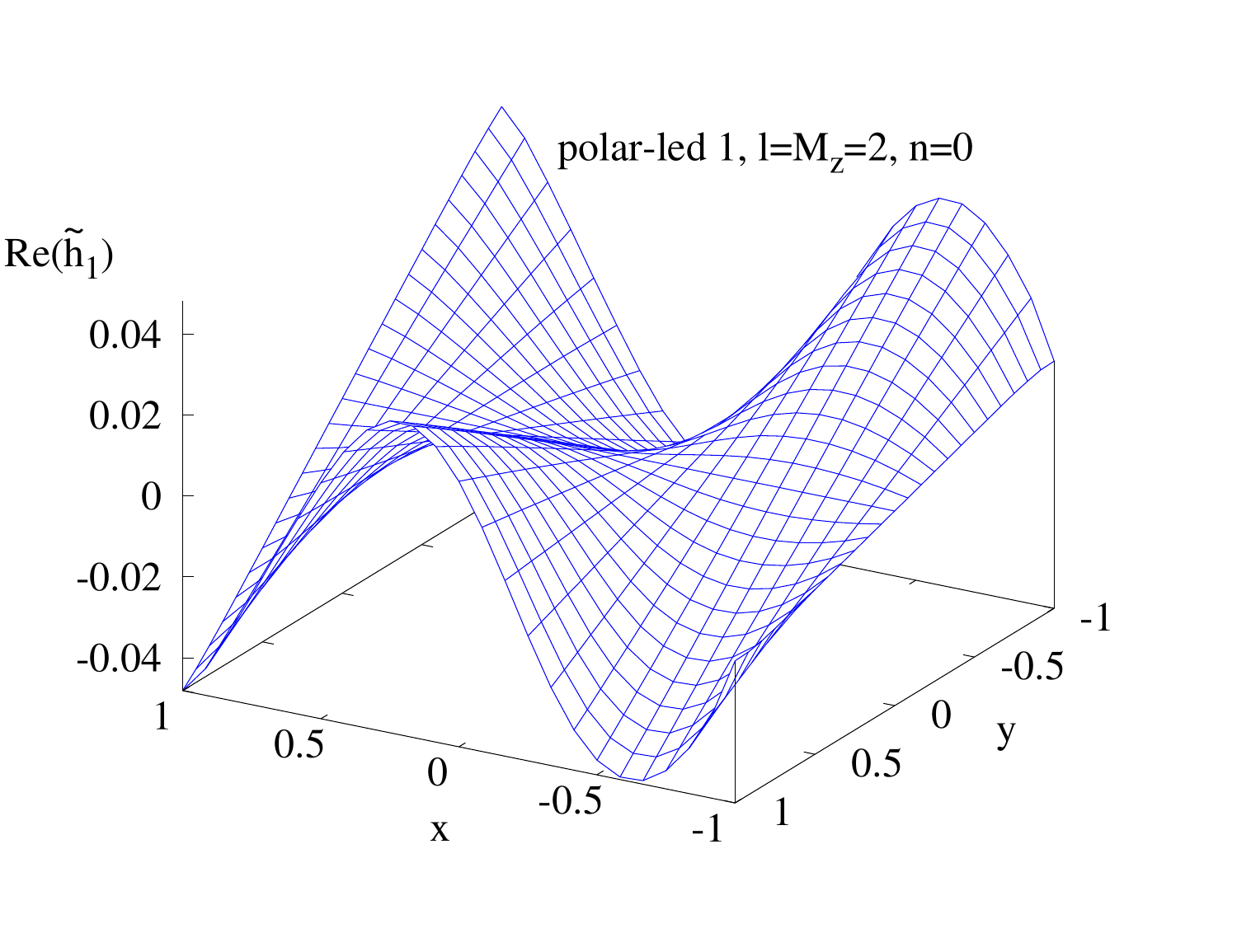}
    \includegraphics[trim=40 0 50 50,clip,angle=-90,width=0.32\textwidth]{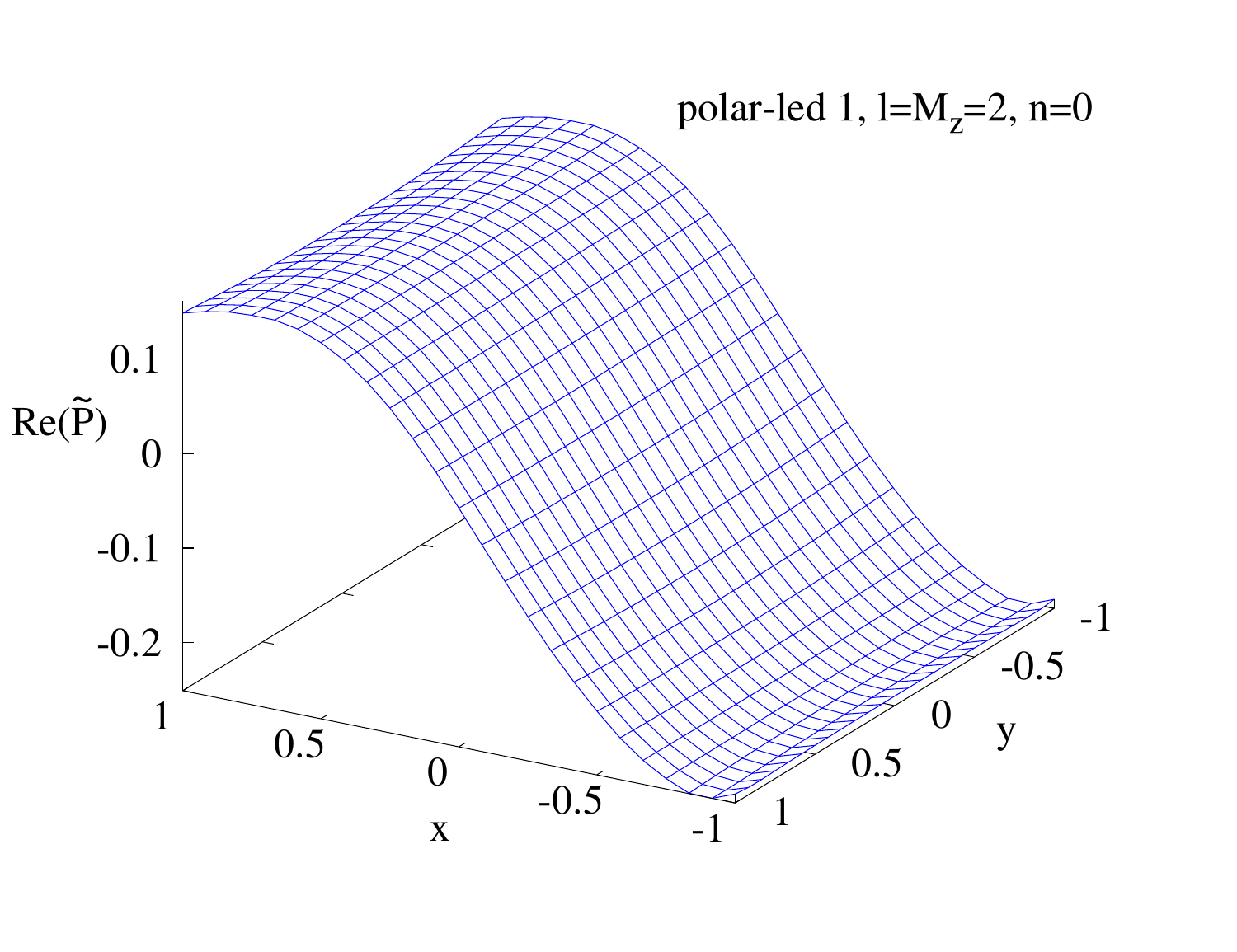}
    \includegraphics[trim=50 0 50 50,clip,angle=-90,width=0.32\textwidth]{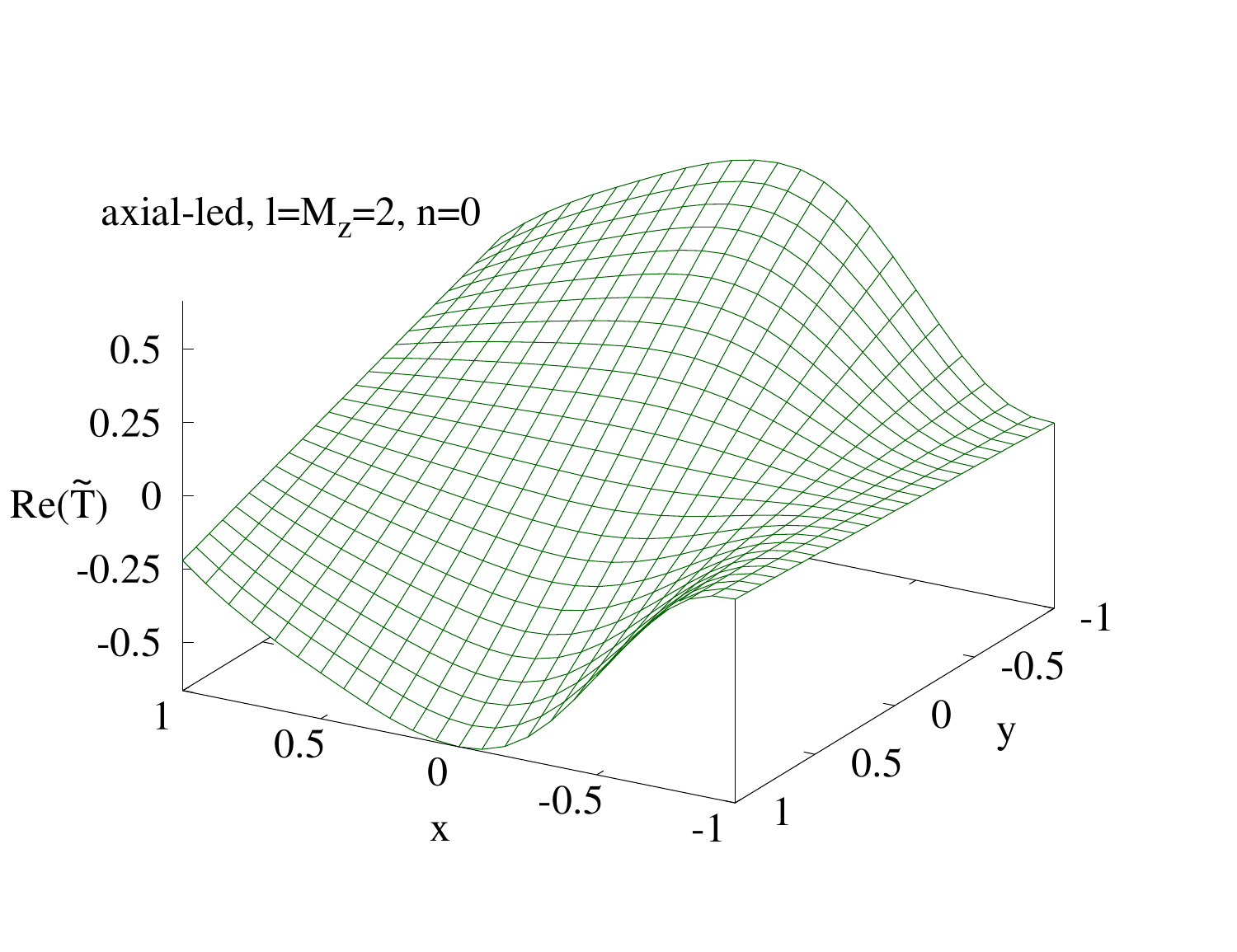}
    \includegraphics[trim=50 0 50 50,clip,angle=-90,width=0.32\textwidth]{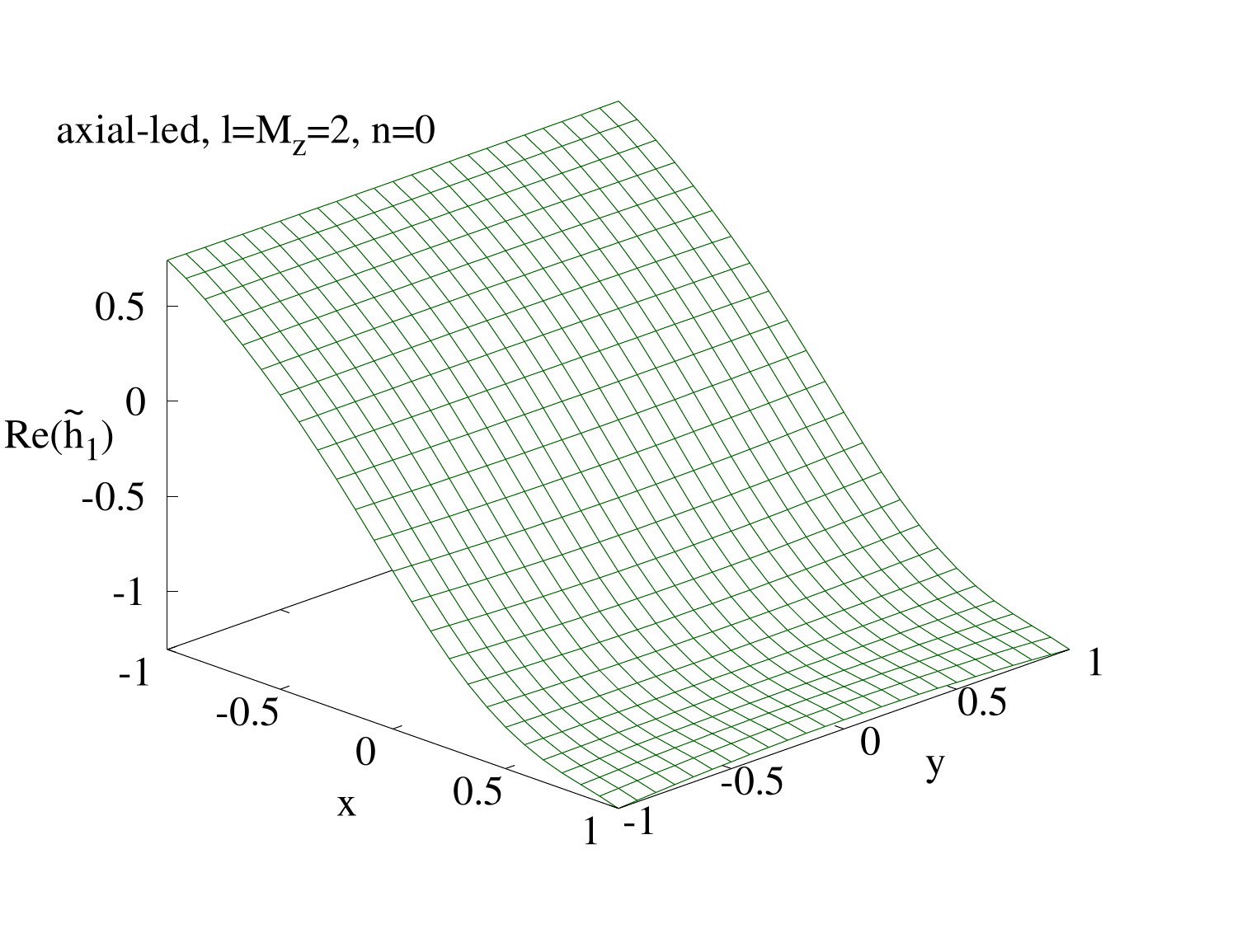}
    \includegraphics[trim=50 0 40 50,clip,angle=-90,width=0.32\textwidth]{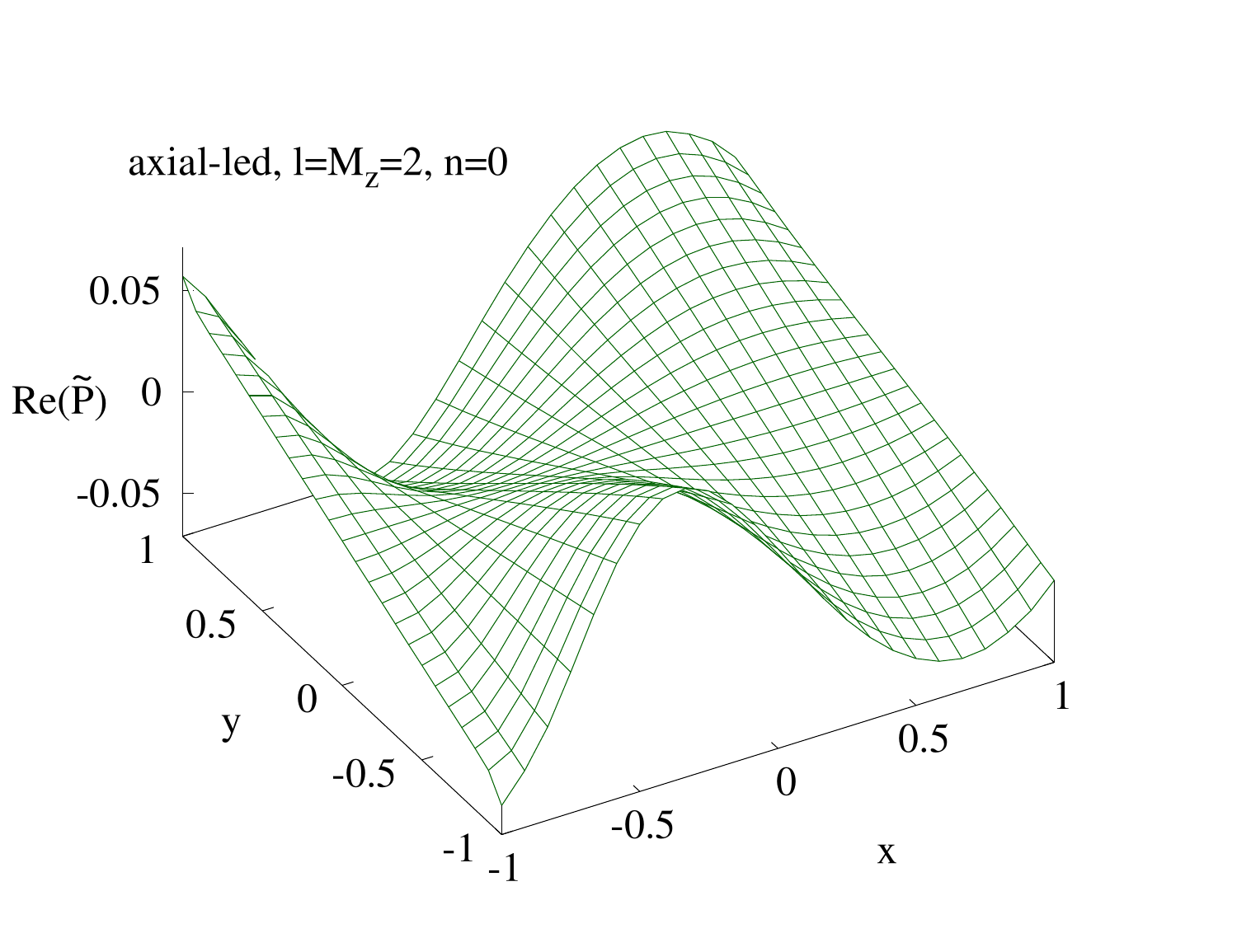}
    \includegraphics[trim=50 0 50 50,clip,angle=-90,width=0.32\textwidth]{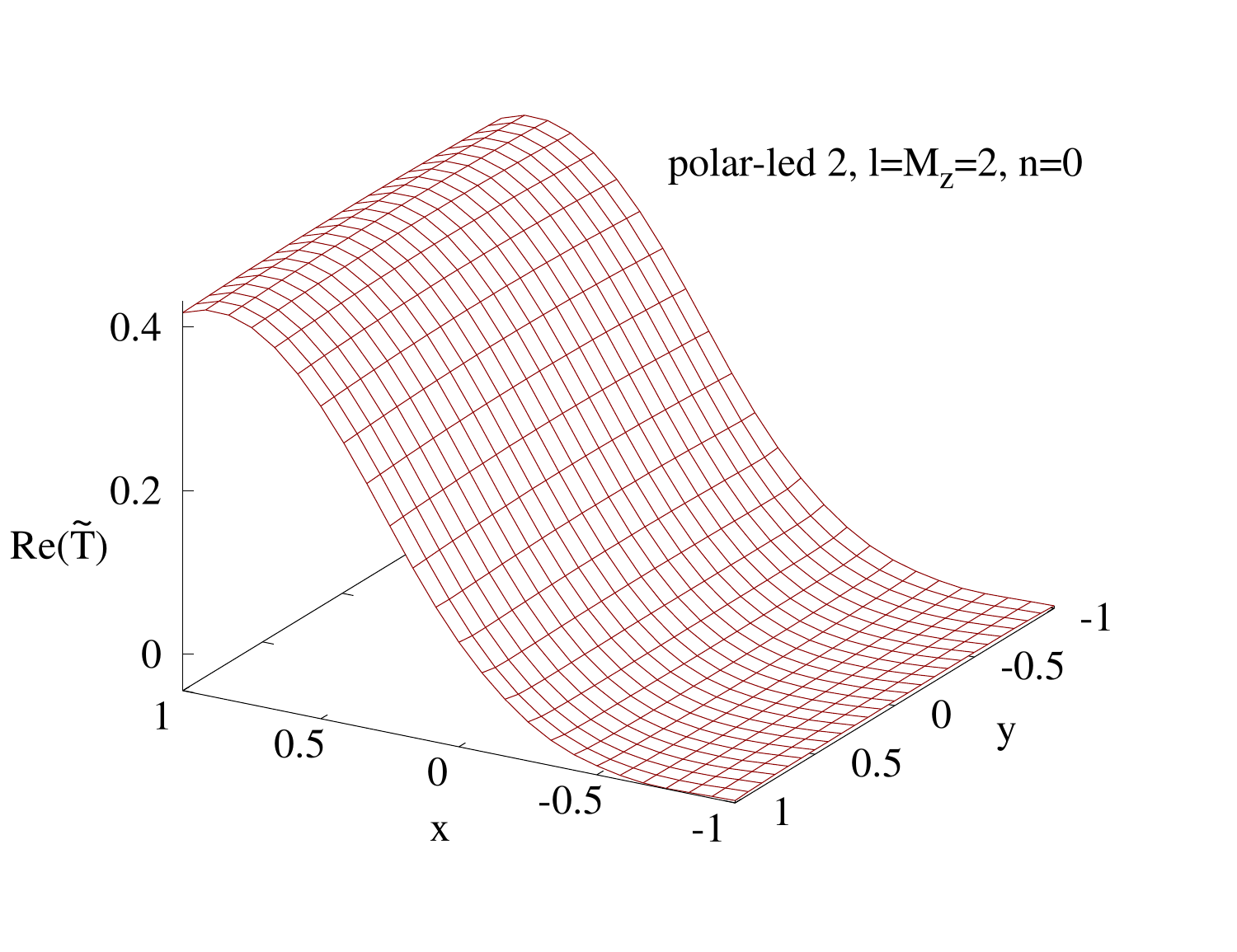}
    \includegraphics[trim=50 0 40 50,clip,angle=-90,width=0.32\textwidth]{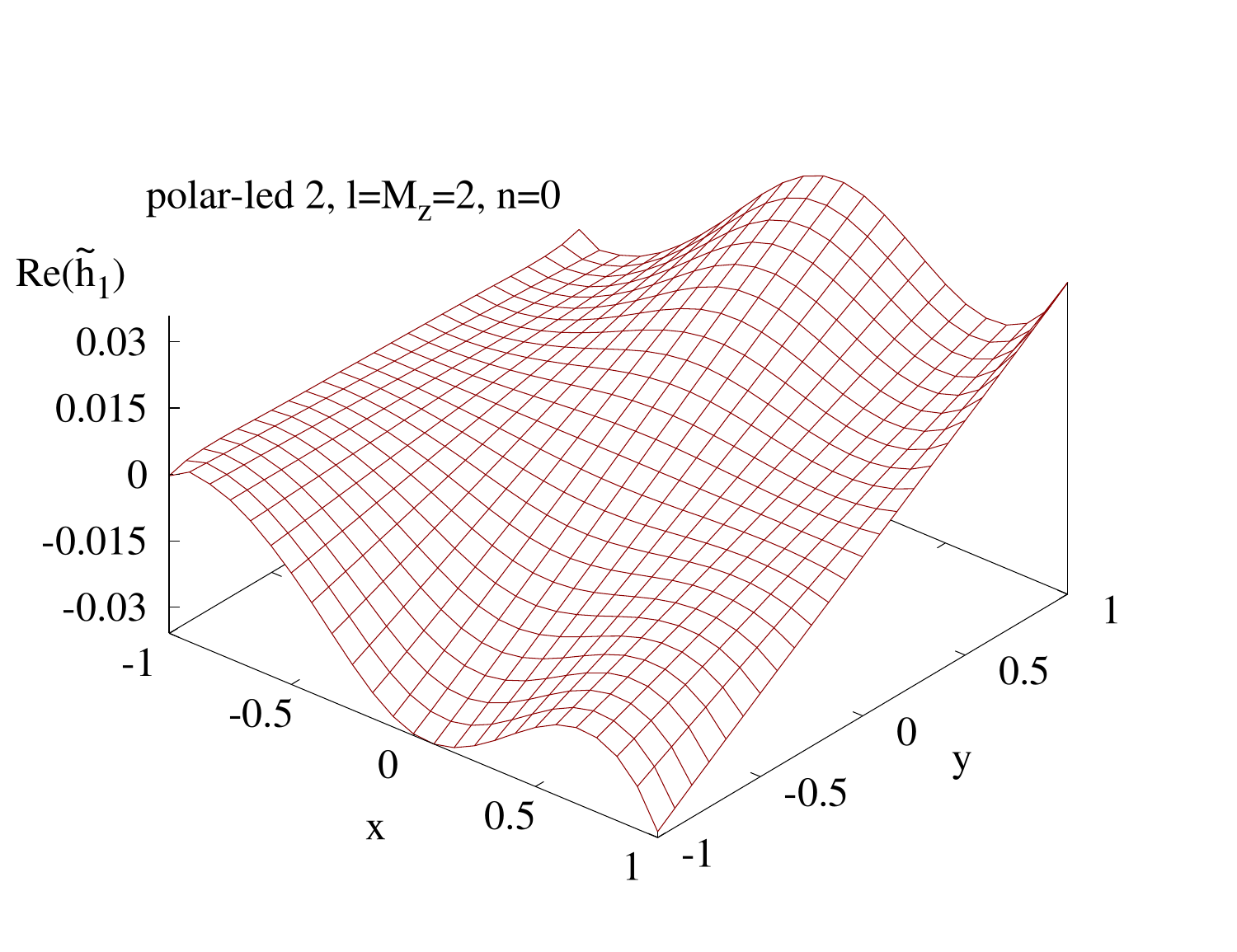}
    \includegraphics[trim=50 0 40 50,clip,angle=-90,width=0.32\textwidth]{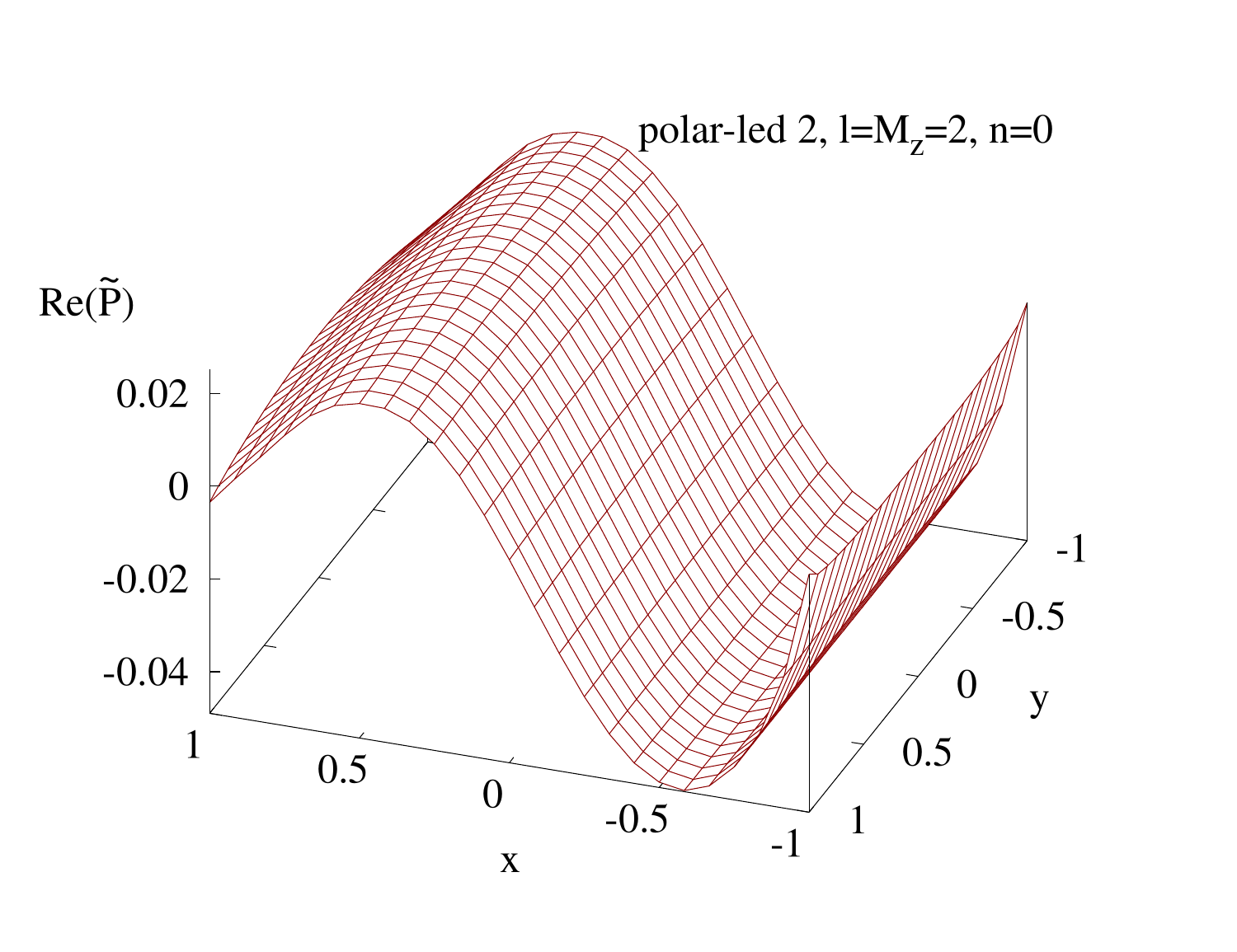}
    \caption{{ Real part of the} perturbation functions $\widetilde T$ (left column), $\widetilde h_1$ (centre column), and $\widetilde P$ (right column) for the fundamental $l=M_z=2$ modes of a wormhole with $r_0=1$ and $w_{-\infty}=0.8$. In blue in the top row are the profiles for the polar-led 1 mode, in green in the second row are for the axial-led mode, and in red in the third row are for the polar-led 2 mode.}
    \label{fig:functions}
\end{figure}

To exemplify how the nature of a mode is imprinted in the perturbation functions, we show in Figure \ref{fig:functions} the typical profiles for the {fundamental ($n=0$)} $l=M_z=2$ modes in a rotating background with $w_{-\infty}=0.8$ and $r_0=1$. In blue we show the functions for the polar-led 1 mode, in green for the axial-led mode, and in red for the polar-led 2 mode. For each mode we exhibit the $\widetilde T$ polar function, $\widetilde h_1$ axial function, and $\widetilde P$ scalar function.  

As these are the perturbation functions of the $(l=2)$-led modes, when a mode is polar-dominated (here in blue and in red), the polar functions $\widetilde T$ and $\widetilde P$ behave predominantly like an even function with respect to the $y$ coordinate, while the axial function $\widetilde h_1$ behaves like an odd function. 
This behaviour reverses when we have an axial-dominated mode (in green). 
Also note that for a polar-dominated mode, the amplitude of the metric polar function is larger than the amplitude of the axial function, and vice versa.

\begin{figure}
    \centering
    \includegraphics[angle=-90,width=0.45\textwidth]{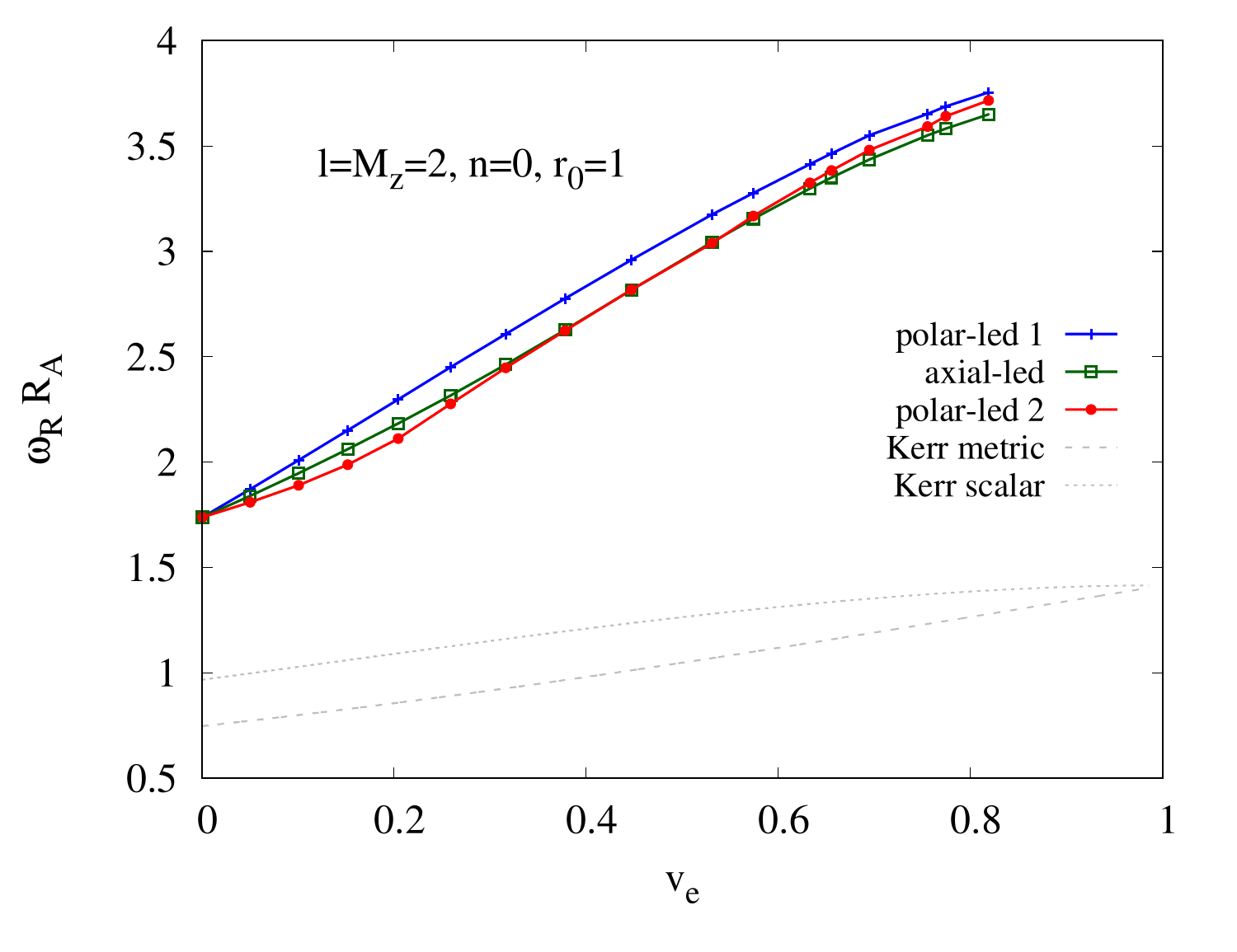}
    \includegraphics[angle=-90,width=0.45\textwidth]{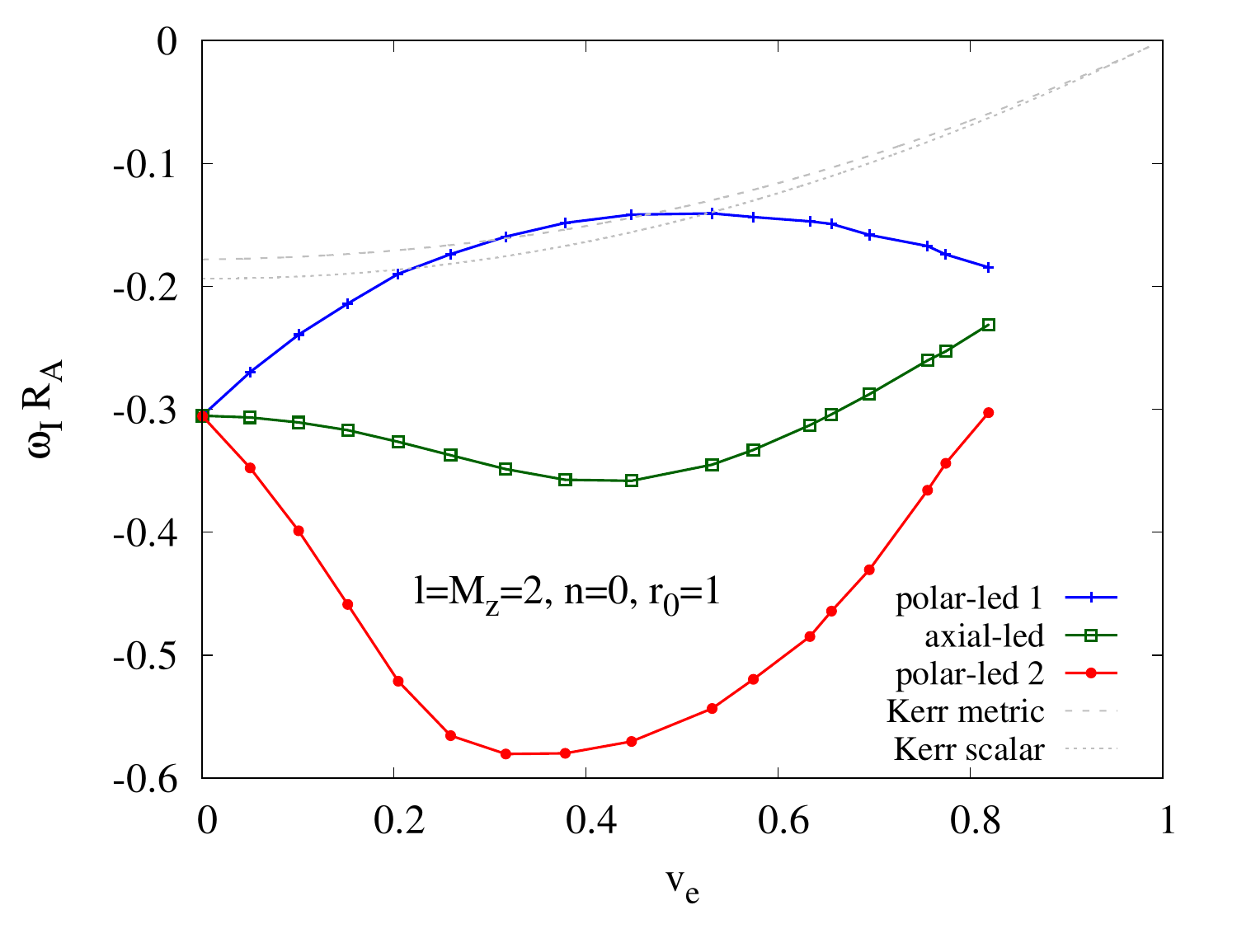}   
    \includegraphics[angle=-90,width=0.45\textwidth]{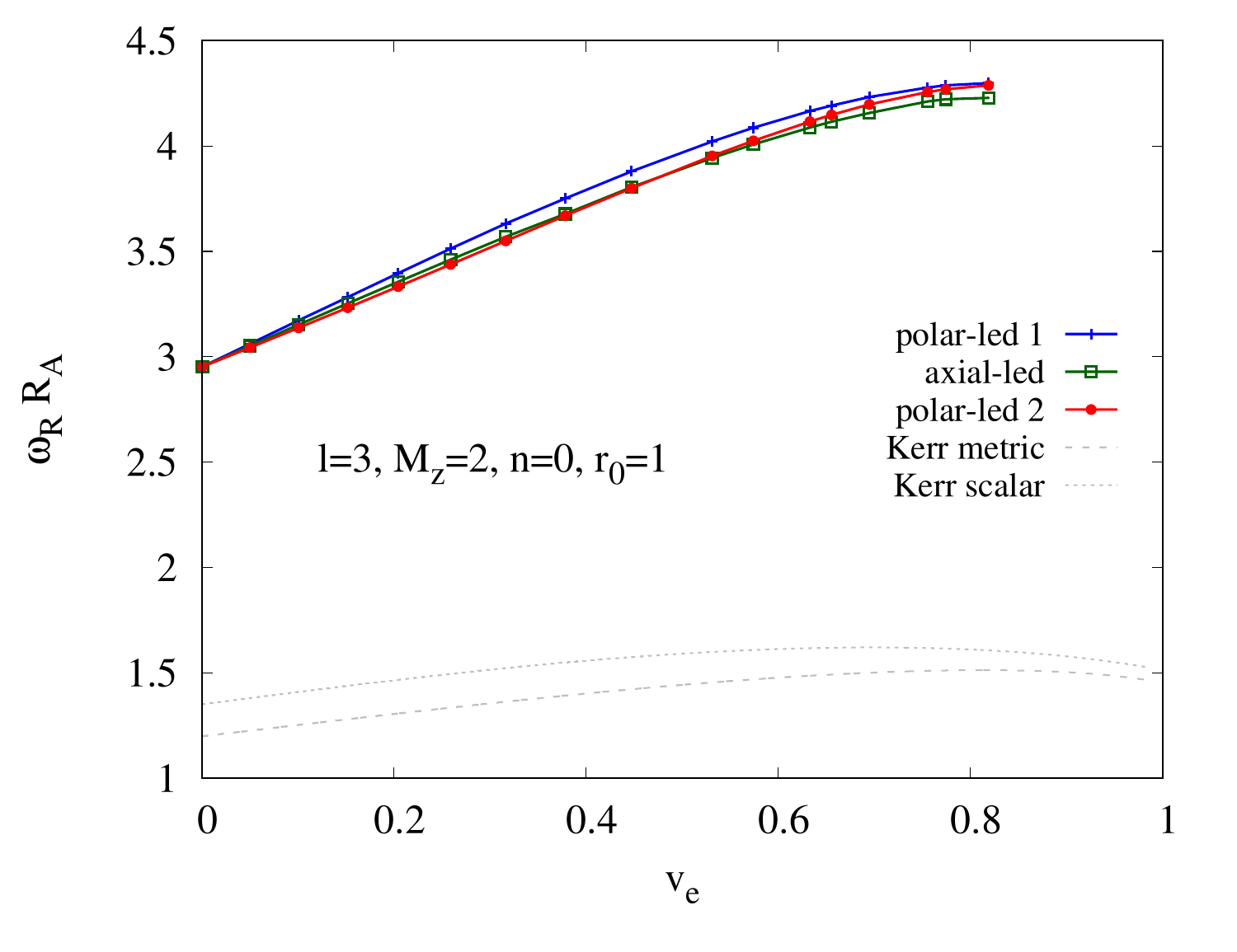}
    \includegraphics[angle=-90,width=0.45\textwidth]{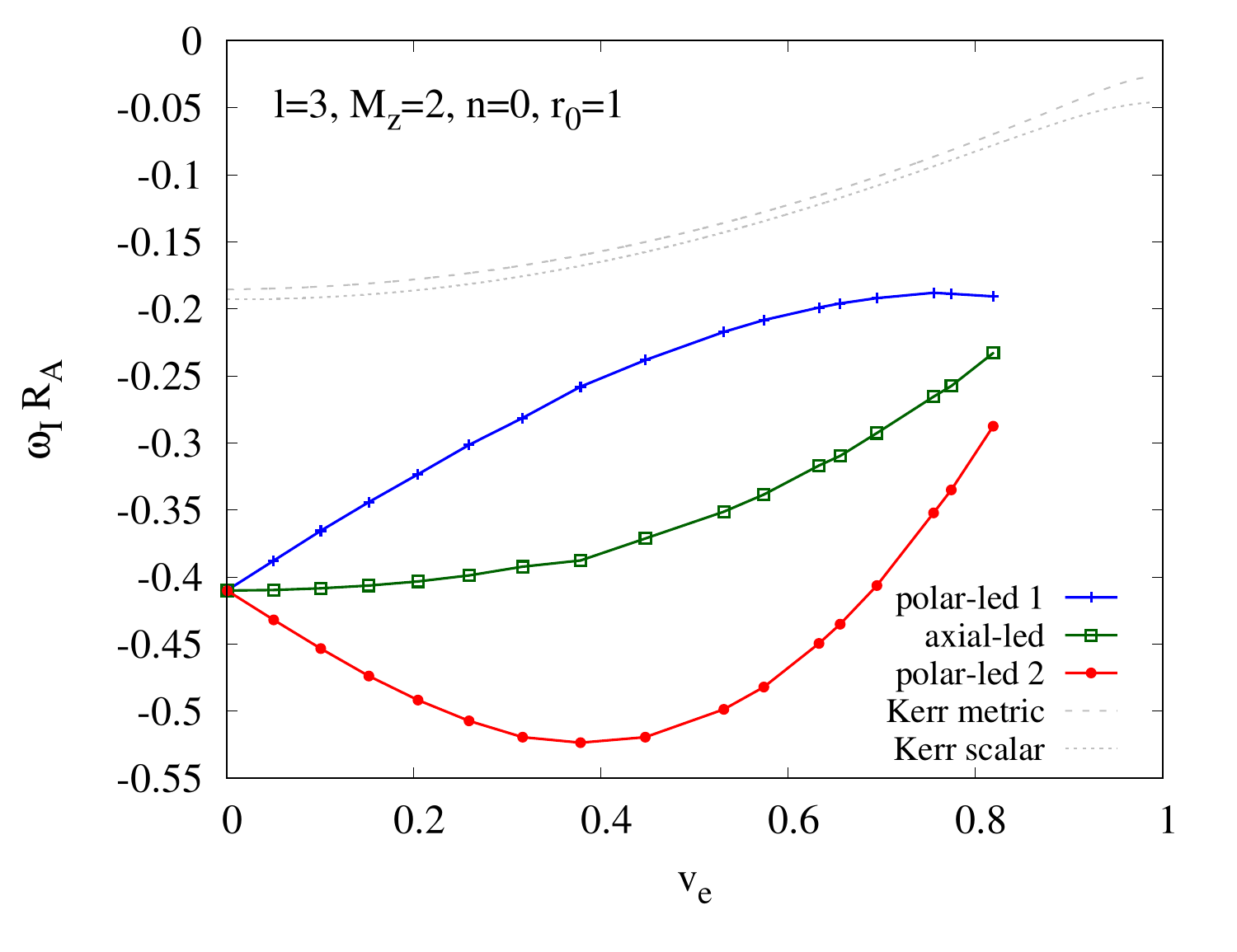}   
    \includegraphics[angle=-90,width=0.45\textwidth]{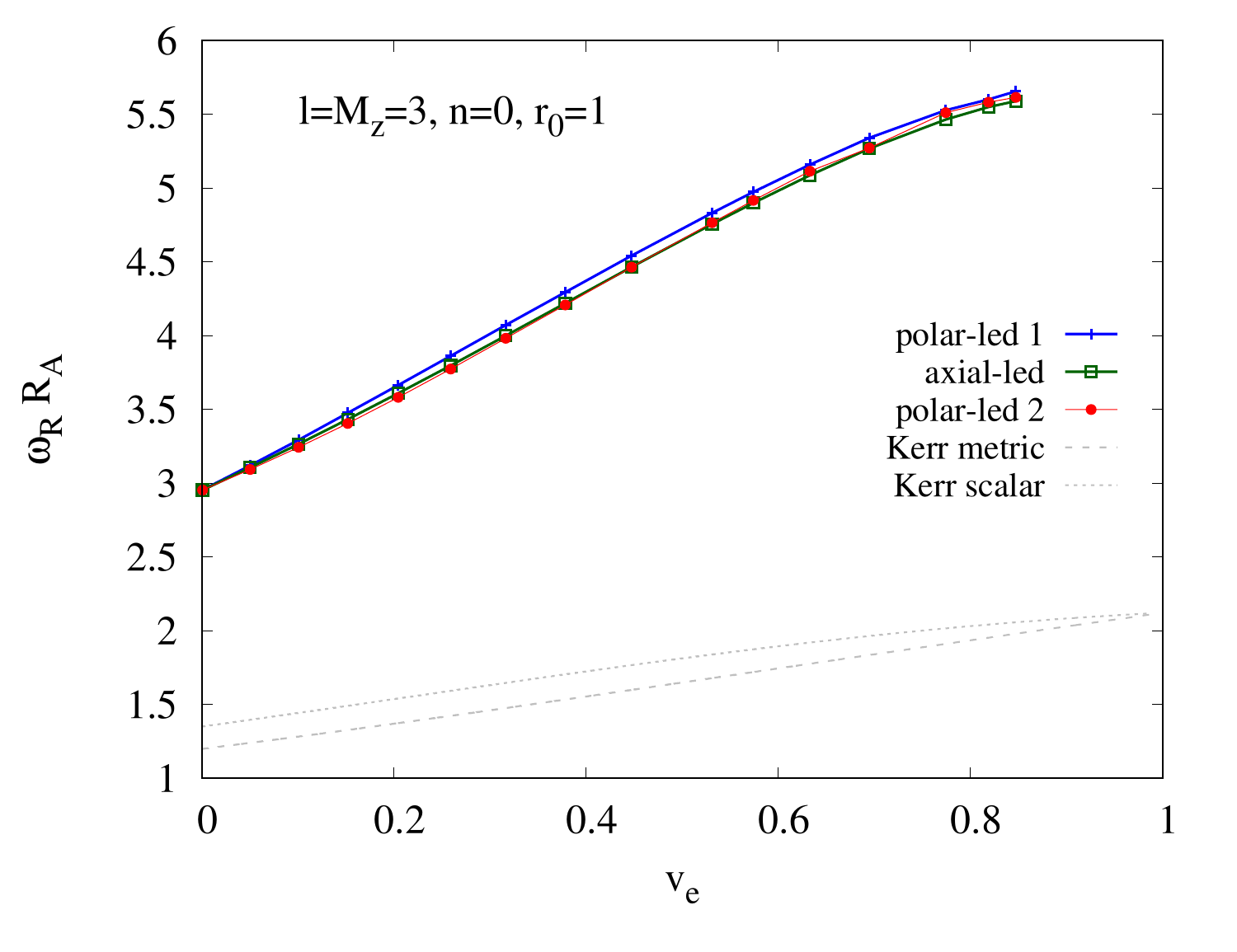}
    \includegraphics[angle=-90,width=0.45\textwidth]{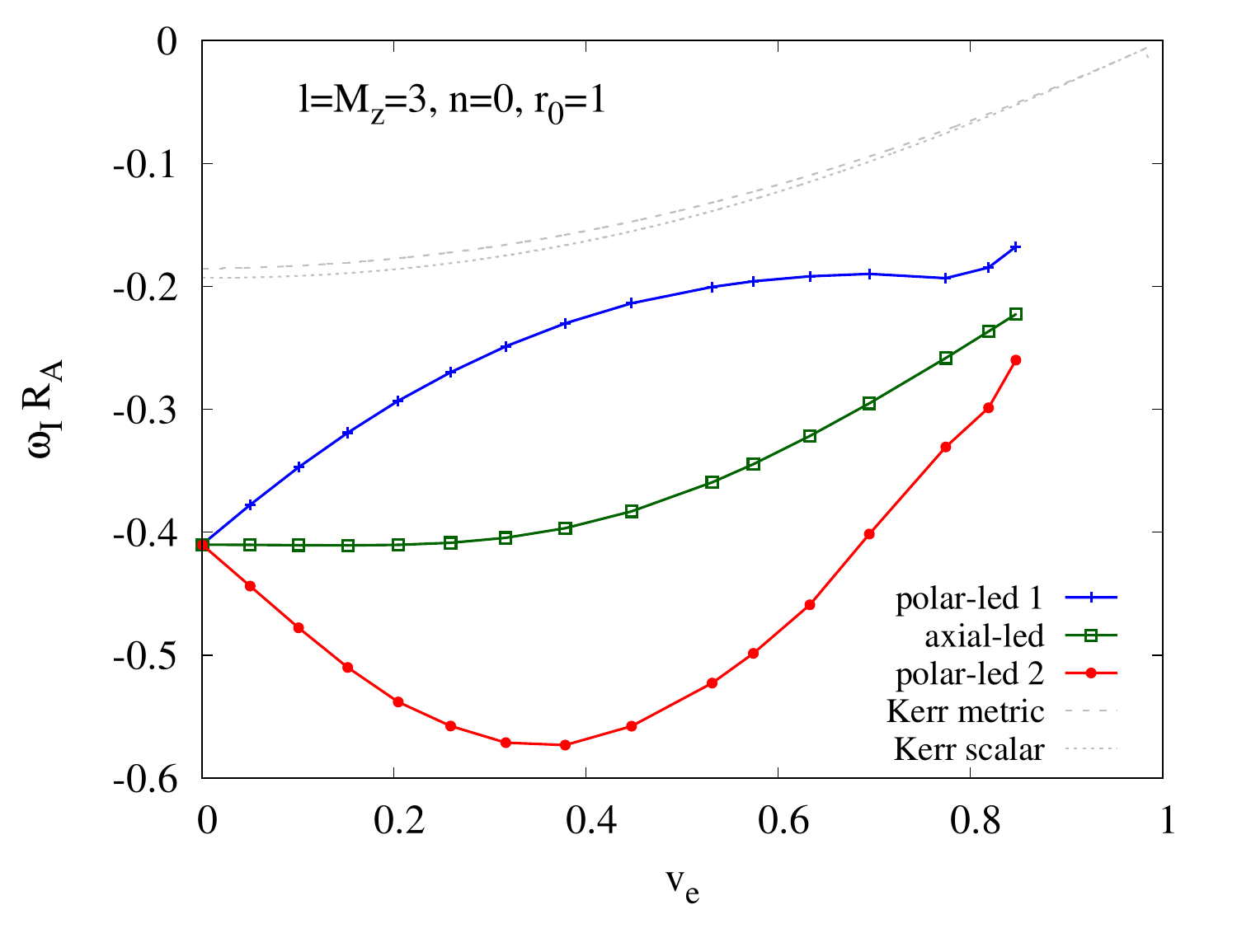}   
    \caption{Fundamental quasinormal modes of rotating Ellis-Bronnikov wormholes and Kerr black holes. In the left column we show the real part 
    and in the right column the imaginary part
    of the frequency scaled with the areal radius $\omega_R \mathrm{R}_A$ as a function of the rotational velocity $\mathrm{v}_e$ of the throat and of the horizon, respectively.}
    \label{fig:ellis_qnms_ve_all}
\end{figure}

In Figure \ref{fig:ellis_qnms_ve_all} we show the complex frequency $\omega$ as a function of the rotational velocity of the throat $\mathrm{v}_e$. In the left column we show the real part of $\omega$ scaled with the areal radius, and in the right column the scaled imaginary part. 
In blue crosses we show the polar-led 1 modes, in green squares the axial-led modes, and in red dots the polar-led 2 modes.

To elucidate the difference in the spectra of wormholes and black holes,
we show in Figure \ref{fig:ellis_qnms_ve_all} the quasinormal modes
of Ellis-Bronnikov wormholes and those of Kerr black holes.
Here Kerr metric 
denotes the metric perturbations of the Kerr spacetime (shown in a dashed grey line),
and Kerr scalar denotes a test scalar field in the Kerr background (shown in a dotted grey line).
In order to make comparisons with the rotating wormholes, we have scaled the Kerr quasinormal modes with respect to the following Kerr quantities in Boyer-Lindquist coordinates.
The rotational velocity of the horizon is
\begin{eqnarray}
    \mathrm{v}_e=\Omega_H \mathrm{R}_e =  \frac{a}{r_H} \, ,
\end{eqnarray}
and the areal radius of the horizon is
\begin{eqnarray}
   \mathrm{R}_A=\sqrt{r_H^2+a^2} \, .
\end{eqnarray}
$M$ is the black hole mass,
$a$ is the Kerr parameter, 
$\Omega_H=a/(r_H^2+a^2)$ is the horizon angular velocity, where $r_H=M+\sqrt{M^2-a^2}$ is the horizon radius, and $\mathrm{R}_e=2M$ is the equatorial radius of the horizon.

In the figure we can appreciate that 
the scaled real part of the modes 
is always larger for Ellis-Bronnikov wormholes than for Kerr black holes.
The absolute value of the scaled imaginary part of the modes is also typically larger 
for the wormholes than for the black holes. The only exception occurs for the $l=M_z=2$ polar-led 1 modes, for a range of $\mathrm{v}_e$ roughly between
$0.2$ and $0.6$, where the
Ellis-Bronnikov modes have a smaller absolute value than the Kerr modes.

In any case, the {triple} breaking of isospectrality of the wormholes due to rotation means that the Ellis-Bronnikov spectrum is richer than the Kerr spectrum. 
Ellis-Bronnikov wormholes present a distinct behaviour in the modes between
even and odd perturbations,
whereas the Kerr modes remain isospectral. In general the Ellis-Bronnikov modes are longer lived than the Kerr modes, especially for the axial-led modes and the second branch of polar-led modes.

Although the precision in our calculations
diminishes as we increase the rotational velocity beyond $\mathrm{v}_e=0.82$, in general we observe a tendency of the modes to come back together as we increase the rotational velocity of the throat. 
This tendency is particularly significant in the imaginary parts of the axial-led modes and the polar-led 2 modes (green and red curves in Figure \ref{fig:ellis_qnms_ve_all}), { especially so for the $l=M_z=3$ modes (bottom panels)}. Hence it is possible that isospectrality is restored as we approach the limiting extremal Kerr black hole.
Note also that in \cite{Richartz:2015saa}, no damped modes were found for $l=M_z=2,3$ for extremal Kerr black holes. 

However, we note that the boundary conditions that we impose in order to obtain the quasinormal modes of the wormholes are different from the ones imposed for black holes: 
We require that the perturbations behave as an outgoing wave at both asymptotic infinities for the wormhole, while for a black hole we require an ingoing wave behaviour at the horizon of a black hole. 
Hence, although the background configuration tends to an extremal Kerr black hole as we increase the rotational velocity, the spectrum of the wormhole does not necessarily tend towards the  spectrum of an extremal black hole.

To end this section we present tables with the numerical values of the quasinormal modes shown in Figure \ref{fig:ellis_qnms_ve_all}. 
In Table \ref{tab:l2m2} we present the values for $l=M_z=2$. The frequencies are scaled with the areal radius $\mathrm{R}_A$ and given as a function of the rotational velocity of the throat $\mathrm{v}_e$. In Tables \ref{tab:l3m2} and \ref{tab:l3m3} we present the results for $l=3, M_z=2$, and $l=M_z=3$ respectively.

\begin{table}
    \centering
    \begin{tabular}{|c||c|c||c|c||c|c|}
\cline{2-7}
\multicolumn{1}{c||}{}
 &	\multicolumn{2}{|c||}{polar-led 1}& \multicolumn{2}{|c||}{axial-led} &
\multicolumn{2}{|c|}{polar-led 2}  
 \\ \hline 
$\mathrm{v}_e$	&	$\omega_R\mathrm{R}_A$	&	$\omega_I\mathrm{R}_A$	&	$\omega_R\mathrm{R}_A$	&	$\omega_I\mathrm{R}_A$	&	$\omega_R\mathrm{R}_A$	&	$\omega_I\mathrm{R}_A$ 	\\\hline \hline 
0.0005	&	1.7390	&	-0.3048	&	1.7387	&	-0.3051	&	1.7384	&	-0.3055	\\ \hline 
0.0501	&	1.8714	&	-0.2696	&	1.8400	&	-0.3065	&	1.8095	&	-0.3476	\\ \hline 
0.1005	&	2.0096	&	-0.2391	&	1.9475	&	-0.3106	&	1.8903	&	-0.3988	\\ \hline 
0.1518	&	2.1512	&	-0.2139	&	2.0614	&	-0.3168	&	1.9881	&	-0.4586	\\ \hline 
0.2043	&	2.2981	&	-0.1898	&	2.1836	&	-0.3264	&	2.1121	&	-0.5211	\\ \hline 
0.2588	&	2.4507	&	-0.1737	&	2.3161	&	-0.3373	&	2.2762	&	-0.5654	\\ \hline 
0.3162	&	2.6083	&	-0.1594	&	2.4628	&	-0.3486	&	2.4473	&	-0.5803	\\ \hline 
0.3781	&	2.7766	&	-0.1482	&	2.6280	&	-0.3572	&	2.6242	&	-0.5798	\\ \hline 
0.4472	&	2.9606	&	-0.1415	&	2.8176	&	-0.3581	&	2.8183	&	-0.5702	\\ \hline 
0.5311	&	3.1753	&	-0.1406	&	3.0440	&	-0.3449	&	3.0395	&	-0.5432	\\ \hline 
0.5739	&	3.2776	&	-0.1434	&	3.1547	&	-0.3329	&	3.1678	&	-0.5195	\\ \hline 
0.6329	&	3.4139	&	-0.1470	&	3.2986	&	-0.3127	&	3.3253	&	-0.4847	\\ \hline 
0.6553	&	3.4642	&	-0.1490	&	3.3503	&	-0.3040	&	3.3842	&	-0.4641	\\ \hline 
0.6948	&	3.5510	&	-0.1580	&	3.4361	&	-0.2876	&	3.4812	&	-0.4304	\\ \hline 
0.7553	&	3.6532	&	-0.1671	&	3.5510	&	-0.2601	&	3.5922	&	-0.3657	\\ \hline 
0.7743	&	3.6882	&	-0.1739	&	3.5828	&	-0.2527	&	3.6411	&	-0.3438	\\ \hline 
0.8189	&	3.7548	&	-0.1844	&	3.6507	&	-0.2310	&	3.7161	&	-0.3025	\\ 
\hline 
    \end{tabular}
    \caption{Fundamental $(l=2)$-led modes with $M_z=2$}
    \label{tab:l2m2}
\end{table}

\begin{table}
    \centering
    \begin{tabular}{|c||c|c||c|c||c|c|}
\cline{2-7}
\multicolumn{1}{c||}{}
 &	\multicolumn{2}{|c||}{polar-led 1}& \multicolumn{2}{|c||}{axial-led} &
\multicolumn{2}{|c|}{polar-led 2}  
 \\ \hline 
$\mathrm{v}_e$	&	$\omega_R\mathrm{R}_A$	&	$\omega_I\mathrm{R}_A$	&	$\omega_R\mathrm{R}_A$	&	$\omega_I\mathrm{R}_A$	&	$\omega_R\mathrm{R}_A$	&	$\omega_I\mathrm{R}_A$ 	\\\hline \hline 
0.0005	&	2.9535	&	-0.4098	&	2.9534	&	-0.4100	&	2.9533	&	-0.4102	\\\hline
0.0501	&	3.0619	&	-0.3878	&	3.0523	&	-0.4096	&	3.0439	&	-0.4320	\\\hline
0.1005	&	3.1725	&	-0.3654	&	3.1524	&	-0.4083	&	3.1372	&	-0.4533	\\\hline
0.1518	&	3.2825	&	-0.3441	&	3.2535	&	-0.4063	&	3.2329	&	-0.4738	\\\hline
0.2043	&	3.3964	&	-0.3233	&	3.3559	&	-0.4033	&	3.3330	&	-0.4917	\\\hline
0.2588	&	3.5132	&	-0.3014	&	3.4609	&	-0.3988	&	3.4382	&	-0.5072	\\\hline
0.3162	&	3.6325	&	-0.2813	&	3.5696	&	-0.3923	&	3.5498	&	-0.5194	\\\hline
0.3781	&	3.7509	&	-0.2581	&	3.6776	&	-0.3876	&	3.6692	&	-0.5235	\\\hline
0.4472	&	3.8800	&	-0.2381	&	3.8050	&	-0.3713	&	3.8002	&	-0.5194	\\\hline
0.5311	&	4.0218	&	-0.2172	&	3.9434	&	-0.3513	&	3.9530	&	-0.4986	\\\hline
0.5739	&	4.0869	&	-0.2084	&	4.0081	&	-0.3385	&	4.0245	&	-0.4820	\\\hline
0.6329	&	4.1656	&	-0.1990	&	4.0880	&	-0.3168	&	4.1162	&	-0.4495	\\\hline
0.6553	&	4.1922	&	-0.1959	&	4.1157	&	-0.3096	&	4.1478	&	-0.4351	\\\hline
0.6948	&	4.2329	&	-0.1919	&	4.1573	&	-0.2926	&	4.1975	&	-0.4063	\\\hline
0.7553	&	4.2776	&	-0.1880	&	4.2113	&	-0.2654	&	4.2563	&	-0.3522	\\\hline
0.7743	&	4.2884	&	-0.1888	&	4.2221	&	-0.2575	&	4.2693	&	-0.3350	\\\hline
0.8189	&	4.2983	&	-0.1907	&	4.2285	&	-0.2329	&	4.2880	&	-0.2875	\\
\hline
    \end{tabular}
    \caption{Fundamental $(l=3)$-led modes with $M_z=2$}
    \label{tab:l3m2}
\end{table}

\begin{table}
    \centering
    \begin{tabular}{|c||c|c||c|c||c|c|}
\cline{2-7}
\multicolumn{1}{c||}{}
 &	\multicolumn{2}{|c||}{polar-led 1}& \multicolumn{2}{|c||}{axial-led} &
\multicolumn{2}{|c|}{polar-led 2}  
 \\ \hline 
$\mathrm{v}_e$	&	$\omega_R\mathrm{R}_A$	&	$\omega_I\mathrm{R}_A$	&	$\omega_R\mathrm{R}_A$	&	$\omega_I\mathrm{R}_A$	&	$\omega_R\mathrm{R}_A$	&	$\omega_I\mathrm{R}_A$ 	\\\hline \hline 
0.0005	&	2.9541	&	-0.4097	&	2.9539	&	-0.4100	&	2.9538	&	-0.4103	\\\hline
0.0501	&	3.1196	&	-0.3776	&	3.1054	&	-0.4101	&	3.0929	&	-0.4437	\\\hline
0.1005	&	3.2939	&	-0.3471	&	3.2652	&	-0.4104	&	3.2432	&	-0.4776	\\\hline
0.1518	&	3.4756	&	-0.3189	&	3.4328	&	-0.4106	&	3.4055	&	-0.5099	\\\hline
0.2043	&	3.6650	&	-0.2930	&	3.6098	&	-0.4102	&	3.5817	&	-0.5380	\\\hline
0.2588	&	3.8630	&	-0.2697	&	3.7980	&	-0.4084	&	3.7739	&	-0.5575	\\\hline
0.3162	&	4.0720	&	-0.2486	&	4.0000	&	-0.4044	&	3.9820	&	-0.5711	\\\hline
0.3781	&	4.2948	&	-0.2300	&	4.2200	&	-0.3966	&	4.2085	&	-0.5731	\\\hline
0.4472	&	4.5429	&	-0.2137	&	4.4657	&	-0.3830	&	4.4641	&	-0.5577	\\\hline
0.5311	&	4.8325	&	-0.2004	&	4.7567	&	-0.3593	&	4.7664	&	-0.5227	\\\hline
0.5739	&	4.9740	&	-0.1957	&	4.9000	&	-0.3445	&	4.9170	&	-0.4984	\\\hline
0.6329	&	5.1591	&	-0.1916	&	5.0876	&	-0.3216	&	5.1156	&	-0.4589	\\\hline
0.6948	&	5.3407	&	-0.1897	&	5.2687	&	-0.2950	&	5.2724	&	-0.4014	\\\hline
0.7743	&	5.5281	&	-0.1932	&	5.4660	&	-0.2582	&	5.5117	&	-0.3305	\\\hline
{0.8189}	&	5.6017	&	-0.1845	&	5.5508	&	-0.2365	&	5.5831	&	-0.2986	\\\hline
{0.8473}	&	5.6566	&	-0.1676	&	5.5906	&	-0.2225	&	5.6163	&	-0.2598	\\
\hline
    \end{tabular}
    \caption{Fundamental $(l=3)$-led modes with $M_z=3$}
    \label{tab:l3m3}
\end{table}

\section{Conclusions}

We have investigated the quasinormal modes of symmetric rapidly rotating Ellis-Bronnikov wormholes.
In order to accomplish this we made use of our recently developed numerical scheme based on a spectral decomposition, tested by determining the modes of Kerr black holes \cite{Blazquez-Salcedo:2023hwg}.
The scheme provides, in principle, the full spectrum of modes for a given background solution and azimuthal number $M_z$.

The quasinormal modes obtained with this scheme can still be classified by the angular number $l$ associated with the mode in the static limit.
However, since the presence of rotation leads to a mixing of the different values of $l$, the resulting modes are now classified as $l$-led modes.
The scheme then produces for the various $l$-led modes both the fundamental modes and their excitations, although in this paper we have focused on the fundamental ones.

By inspecting the perturbation functions, one can further classify the modes.
They can be polar-led or axial-led, depending on the dominating functions.
For the case of rapidly rotating Ellis-Bronnikov wormholes, there are two branches of polar-led modes and one branch of axial-led modes.
Interestingly, in the static limit all three modes coincide for symmetric Ellis-Bronnikov wormholes \cite{Azad:2022qqn}.
However, as shown above, the presence of rotation breaks this triple isospectrality. 

We have demonstrated our scheme for the fundamental $M_z=2$ modes, and then focused our detailed study on the $l=M_z=2$ modes, the $l=3,M_z=2$ modes, and the $l=M_z=3$ modes, which might be the most relevant from an observational point of view.
For those modes we have determined the axial-led branch and the two polar-led branches in the range $0<\mathrm{v}_e < {0.85}$, where $\mathrm{v}_e$ is the rotational velocity of the throat of the background solutions.
We recall that the range of $\mathrm{v}_e$ is limited by $\mathrm{v}_e=1$, since in this limit the background solutions tend to the extremal Kerr black hole solution.

When comparing the modes of the wormholes with those of Kerr black holes we note significant differences.
For finite (non-extremal) rotational velocity of the throat all three wormhole branches are distinct in contrast to Kerr black holes.
In the extremal limit $\mathrm{v}_e \to 1$ the imaginary part of the axial-led and the polar-led 2 wormhole modes might possibly tend to the corresponding Kerr values.
However, we do not observe a similar tendency for the real part of the modes.
To confirm these tendencies, we will need to extend our studies beyond $\mathrm{v}_e>0.8$. 
In order to do so with sufficient accuracy in the calculations, it may be necessary to adopt another parametrization of the rotating wormhole background \cite{Volkov:2021blw}.

We have not observed any unstable modes in our analysis of $(l\ge 2)$-led modes of the Ellis-Bronnikov wormholes.
However, the static Ellis-Bronnikov wormholes are known to possess an unstable radial mode \cite{Shinkai:2002gv,Gonzalez:2008wd,Gonzalez:2008xk,Cremona:2018wkj}.
While there is first evidence, that these wormholes might be stabilized by rotation \cite{Azad:2023iju}, it still remains a challenge to confirm (or refute) this appealing possibility by performing a complete linear stability analysis in the rotating case.

\section*{Acknowledgement}
We gratefully acknowledge support by DAAD, DFG project Ku612/18-1, 
FCT project PTDC/FIS-AST/3041/2020, 
and MICINN project PID2021-125617NB-I00 ``QuasiMode".
JLBS gratefully acknowledges support from Santander-UCM project PR44/21‐29910.


\begin{thebibliography}{99}

\bibitem{Morris:1988cz}
M.~S.~Morris and K.~S.~Thorne,
``Wormholes in space-time and their use for interstellar travel: A tool for teaching general relativity,''
Am. J. Phys. \textbf{56}, 395-412 (1988)

\bibitem{Visser:1995cc} 
  M.~Visser,
  ``Lorentzian wormholes: From Einstein to Hawking,''
  Woodbury, USA: AIP (1995).


\bibitem{Lobo:2017}
F.~S.~N.~Lobo, 
``Wormholes, Warp Drives and Energy Conditions,'' 
Springer Cham (2017).


\bibitem{Ellis:1973yv} 
  H.~G.~Ellis,
  J.\ Math.\ Phys.\  {\bf 14}, 104 (1973)

\bibitem{Bronnikov:1973fh} 
  K.~A.~Bronnikov,
  Acta Phys.\ Polon.\ B {\bf 4}, 251 (1973)

\bibitem{Ellis:1979bh} 
  H.~G.~Ellis,
  Gen.\ Rel.\ Grav.\  {\bf 10}, 105 (1979)

\bibitem{Blazquez-Salcedo:2020czn}
J.~L.~Bl\'azquez-Salcedo, C.~Knoll and E.~Radu,
Phys. Rev. Lett. \textbf{126}, 101102 (2021)

\bibitem{Konoplya:2021hsm}
R.~A.~Konoplya and A.~Zhidenko,
Phys. Rev. Lett. \textbf{128}, 091104 (2022)

\bibitem{Blazquez-Salcedo:2021udn}
J.~L.~Bl\'azquez-Salcedo, C.~Knoll and E.~Radu,
Eur. Phys. J. C \textbf{82}, 533 (2022)

\bibitem{Kanti:2011jz}
  P.~Kanti, B.~Kleihaus and J.~Kunz,
  Phys.\ Rev.\ Lett.\  {\bf 107}, 271101 (2011)

\bibitem{Kanti:2011yv}
  P.~Kanti, B.~Kleihaus and J.~Kunz,
  Phys.\ Rev.\ D {\bf 85}, 044007 (2012).

\bibitem{Antoniou:2019awm}
G.~Antoniou, A.~Bakopoulos, P.~Kanti, B.~Kleihaus and J.~Kunz,
Phys. Rev. D \textbf{101}, 024033 (2020)

\bibitem{Bakopoulos:2021liw}
A.~Bakopoulos, C.~Charmousis and P.~Kanti,
JCAP \textbf{05}, 022 (2022)

\bibitem{Cramer:1994qj}
  J.~G.~Cramer, R.~L.~Forward, M.~S.~Morris, M.~Visser, G.~Benford and G.~A.~Landis,
  Phys.\ Rev.\ D {\bf 51}, 3117 (1995)

\bibitem{Safonova:2001vz}
  M.~Safonova, D.~F.~Torres and G.~E.~Romero,
  Phys.\ Rev.\ D {\bf 65}, 023001 (2002)

\bibitem{Perlick:2003vg}
  V.~Perlick,
  Phys.\ Rev.\ D {\bf 69}, 064017 (2004)

\bibitem{Nandi:2006ds}
  K.~K.~Nandi, Y.~Z.~Zhang and A.~V.~Zakharov,
  Phys.\ Rev.\ D {\bf 74}, 024020 (2006)

\bibitem{Abe:2010ap}
  F.~Abe,
  Astrophys.\ J.\  {\bf 725}, 787 (2010)

\bibitem{Toki:2011zu}
  Y.~Toki, T.~Kitamura, H.~Asada and F.~Abe,
  Astrophys.\ J.\  {\bf 740}, 121 (2011)

\bibitem{Nakajima:2012pu}
  K.~Nakajima and H.~Asada,
  Phys.\ Rev.\ D {\bf 85}, 107501 (2012)

\bibitem{Tsukamoto:2012xs}
  N.~Tsukamoto, T.~Harada and K.~Yajima,
  Phys.\ Rev.\ D {\bf 86}, 104062 (2012)

\bibitem{Kuhfittig:2013hva}
  P.~K.~F.~Kuhfittig,
  Eur. Phys.\ J. \ C {\bf 74},  2818 (2014)
 
\bibitem{Bambi:2013nla}
  C.~Bambi,
  Phys.\ Rev.\ D {\bf 87}, 107501 (2013)

\bibitem{Takahashi:2013jqa}
  R.~Takahashi and H.~Asada,
  Astrophys.\ J.\  {\bf 768}, L16 (2013)

\bibitem{Tsukamoto:2016zdu}
  N.~Tsukamoto and T.~Harada,
  Phys.\ Rev.\ D {\bf 95},   024030 (2017)

\bibitem{Nedkova:2013msa}
  P.~G.~Nedkova, V.~K.~Tinchev and S.~S.~Yazadjiev,
  Phys.\ Rev.\ D {\bf 88},   124019 (2013)

\bibitem{Ohgami:2015nra}
  T.~Ohgami and N.~Sakai,
  Phys.\ Rev.\ D {\bf 91}, 124020 (2015)

\bibitem{Shaikh:2018kfv}
  R.~Shaikh,
  Phys.\ Rev.\ D {\bf 98},  024044 (2018)
 

\bibitem{Gyulchev:2018fmd}
  G.~Gyulchev, P.~Nedkova, V.~Tinchev and S.~Yazadjiev,
  Eur.\ Phys.\ J.\ C {\bf 78} , 544 (2018).


\bibitem{Bouhmadi-Lopez:2021zwt}
M.~Bouhmadi-L\'opez, C.~Y.~Chen, X.~Y.~Chew, Y.~C.~Ong and D.~h.~Yeom,
JCAP \textbf{10}, 059 (2021)

\bibitem{Guerrero:2022qkh}
M.~Guerrero, G.~J.~Olmo, D.~Rubiera-Garc\'ia and D.~G\'omez S\'aez-Chill\'on,
Phys. Rev. D \textbf{105}, 084057 (2022)


\bibitem{Huang:2023yqd}
H.~Huang, J.~Kunz, J.~Yang and C.~Zhang,
Phys. Rev. D \textbf{107}, 104060 (2023)

\bibitem{Harko:2008vy}
  T.~Harko, Z.~Kovacs and F.~S.~N.~Lobo,
  Phys.\ Rev.\ D {\bf 78}, 084005 (2008)

\bibitem{Harko:2009xf}
  T.~Harko, Z.~Kovacs and F.~S.~N.~Lobo,
  Phys.\ Rev.\ D {\bf 79}, 064001 (2009)

\bibitem{Bambi:2013jda}
  C.~Bambi,
  Phys.\ Rev.\ D {\bf 87}, 084039 (2013)

\bibitem{Zhou:2016koy}
  M.~Zhou, A.~Cardenas-Avendano, C.~Bambi, B.~Kleihaus and J.~Kunz,
  Phys.\ Rev.\ D {\bf 94}, 024036 (2016)

\bibitem{Lamy:2018zvj}
  F.~Lamy, E.~Gourgoulhon, T.~Paumard and F.~H.~Vincent,
  Class.\ Quant.\ Grav.\  {\bf 35},   115009 (2018)

\bibitem{Deligianni:2021ecz}
E.~Deligianni, J.~Kunz, P.~Nedkova, S.~Yazadjiev and R.~Zheleva,
Phys. Rev. D \textbf{104},  024048 (2021)

\bibitem{Deligianni:2021hwt}
E.~Deligianni, B.~Kleihaus, J.~Kunz, P.~Nedkova and S.~Yazadjiev,
Phys. Rev. D \textbf{104},  064043 (2021)

\bibitem{Damour:2007ap}
T.~Damour and S.~N.~Solodukhin,
Phys. Rev. D \textbf{76}, 024016 (2007)

\bibitem{LIGOScientific:2016aoc}
B.~P.~Abbott \textit{et al.} [LIGO Scientific and Virgo],
Phys. Rev. Lett. \textbf{116}, 061102 (2016).

\bibitem{Konoplya:2005et}
  R.~A.~Konoplya and C.~Molina,
  Phys.\ Rev.\ D {\bf 71}, 124009 (2005)

\bibitem{Kim:2008zzj}
  S.~W.~Kim,
  Prog.\ Theor.\ Phys.\ Suppl.\  {\bf 172}, 21 (2008)

\bibitem{Konoplya:2010kv}
  R.~A.~Konoplya and A.~Zhidenko,
  Phys.\ Rev.\ D {\bf 81}, 124036 (2010)

\bibitem{Konoplya:2016hmd}
  R.~A.~Konoplya and A.~Zhidenko,
  JCAP {\bf 1612}, 043 (2016)
  
\bibitem{Volkel:2018hwb}
  S.~H.~V\"olkel and K.~D.~Kokkotas,
  Class.\ Quant.\ Grav.\  {\bf 35}, 105018 (2018)

\bibitem{Aneesh:2018hlp}
S.~Aneesh, S.~Bose and S.~Kar,
Phys. Rev. D \textbf{97}, 124004 (2018)

\bibitem{Konoplya:2018ala}
R.~A.~Konoplya,
Phys. Lett. B \textbf{784}, 43 (2018)

\bibitem{Blazquez-Salcedo:2018ipc}
J.~L.~Bl\'azquez-Salcedo, X.~Y.~Chew and J.~Kunz,
Phys. Rev. D \textbf{98}, 044035 (2018)

\bibitem{Konoplya:2019hml}
R.~A.~Konoplya, A.~F.~Zinhailo and Z.~Stuchl\'\i{}k,
Phys. Rev. D \textbf{99}, 124042 (2019)

\bibitem{Churilova:2019qph}
M.~S.~Churilova, R.~A.~Konoplya and A.~Zhidenko,
Phys. Lett. B \textbf{802}, 135207 (2020)

\bibitem{Jusufi:2020mmy}
K.~Jusufi,
Gen. Rel. Grav. \textbf{53}, 87 (2021)

\bibitem{Bronnikov:2021liv}
K.~A.~Bronnikov, R.~A.~Konoplya and T.~D.~Pappas,
Phys. Rev. D \textbf{103}, 124062 (2021)

\bibitem{Gonzalez:2022ote}
P.~A.~Gonz\'alez, E.~Papantonopoulos, \'A.~Rinc\'on and Y.~V\'asquez,
Phys. Rev. D \textbf{106}, 024050 (2022)

\bibitem{Azad:2022qqn}
B.~Azad, J.~L.~Bl\'azquez-Salcedo, X.~Y.~Chew, J.~Kunz and D.~h.~Yeom,
Phys. Rev. D \textbf{107}, 084024 (2023)

\bibitem{Bueno:2017hyj}
P.~Bueno, P.~A.~Cano, F.~Goelen, T.~Hertog and B.~Vercnocke,
Phys. Rev. D \textbf{97}, 024040 (2018)

\bibitem{Kashargin:2007mm}
P.~E.~Kashargin and S.~V.~Sushkov,
Grav. Cosmol. \textbf{14}, 80 (2008)

\bibitem{Kashargin:2008pk}
P.~E.~Kashargin and S.~V.~Sushkov,
Phys. Rev. D \textbf{78}, 064071 (2008)

\bibitem{Kleihaus:2014dla}
B.~Kleihaus and J.~Kunz,
Phys. Rev. D \textbf{90}, 121503 (2014)

\bibitem{Chew:2016epf}
X.~Y.~Chew, B.~Kleihaus and J.~Kunz,
Phys. Rev. D \textbf{94}, no.10, 104031 (2016)

\bibitem{Cisterna:2023uqf}
A.~Cisterna, K.~M\"uller, K.~Pallikaris and A.~Vigan\`o,
Phys. Rev. D \textbf{108}, 024066 (2023)

\bibitem{Blazquez-Salcedo:2023hwg}
J.~L.~Bl\'azquez-Salcedo, F.~S.~Khoo, J.~Kunz and L.~M.~Gonz\'alez-Romero,
[arXiv:2312.10754 [gr-qc]].

\bibitem{Advanpix}
P. Holoborodko, 
Advanpix 5.1.0.15432,
http://www.advanpix.com.

\bibitem{Richartz:2015saa}
M.~Richartz,
Phys. Rev. D \textbf{93}, no.6, 064062 (2016)


\bibitem{Volkov:2021blw}
M.~S.~Volkov,
Phys. Rev. D \textbf{104}, no.12, 124064 (2021)



\bibitem{Shinkai:2002gv}
  H.~a.~Shinkai and S.~A.~Hayward,
  Phys.\ Rev.\ D {\bf 66}, 044005 (2002)

\bibitem{Gonzalez:2008wd}
  J.~A.~Gonz\'alez, F.~S.~Guzman and O.~Sarbach,
  Class.\ Quant.\ Grav.\  {\bf 26}, 015010 (2009)

\bibitem{Gonzalez:2008xk}
  J.~A.~Gonz\'alez, F.~S.~Guzman and O.~Sarbach,
  Class.\ Quant.\ Grav.\  {\bf 26},  015011 (2009)

\bibitem{Cremona:2018wkj}
F.~Cremona, F.~Pirotta and L.~Pizzocchero,
Gen. Rel. Grav. \textbf{51}, 19 (2019)

\bibitem{Azad:2023iju}
B.~Azad, J.~L.~Bl\'azquez-Salcedo, F.~S.~Khoo and J.~Kunz,
Phys. Lett. B \textbf{848}, 138349 (2024)

\end{thebibliography}
\end{document}